\documentclass[12pt,a4paper]{article}
\usepackage{a4wide}
\usepackage{amsmath}
\usepackage{amssymb}
\usepackage{amsfonts}
\usepackage{ifpdf}
\usepackage{graphicx,epsfig}
\usepackage{subfigure}
\usepackage{exscale}
\usepackage{float}
\usepackage{bbm}
\usepackage[numbers,sort&compress]{natbib}

\newcommand{\Z}{{\mathbb{Z}}}
\newcommand{\R}{{\mathbb{R}}}

\newcommand{\RP}{{\mathbb{R}P}}
\newcommand{\CP}{{\mathbb{C}P}}

\makeatletter
\@addtoreset{equation}{section}
\makeatother

\title{\vspace{-1cm} Topological Lattice Actions\footnote{Dedicated to Ferenc Niedermayer on the occasion of his 65th birthday}}

\author{W.\ Bietenholz$^a$, U.\ Gerber$^b$, M.\ Pepe$^c$, and U.-J.\ Wiese$^b$ 
\\ \\
$^a$ Instituto de Ciencias Nucleares \\
Universidad Aut\'{o}noma de M\'{e}xico \\
A.\ P.\ 70-543, C.\ P.\ 04510 Distrito Federal, Mexico \\ \\
$^b$ Albert Einstein Center for Fundamental Physics \\
Institute for Theoretical Physics, Bern University \\
Sidlerstrasse 5, CH-3012 Bern, Switzerland \\ \\
$^c$ INFN, Sezione di Milano-Bicocca \\ 
Edificio U2, Piazza della Scienza 3, 20126 Milano, Italy \\ \\
}

\begin{document} 

\maketitle

\vspace{-1cm}

\begin{abstract} \normalsize

We consider lattice field theories with topological actions, which are invariant
against small deformations of the fields. Some of these actions have infinite 
barriers separating different topological sectors. Topological actions do not 
have the correct classical continuum limit and they cannot be treated using 
perturbation theory, but they still yield the correct quantum continuum limit. 
To show this, we present analytic studies of the 1-d $O(2)$ and $O(3)$ model, 
as well as Monte Carlo simulations of the 2-d $O(3)$ model using topological 
lattice actions. Some topological actions obey and others violate a lattice
Schwarz inequality between the action and the topological charge $Q$. 
Irrespective of this, in the 2-d $O(3)$ model the topological susceptibility 
$\chi_t = \langle Q^2 \rangle/V$ is logarithmically divergent in the continuum 
limit. Still, at non-zero distance the correlator of the topological charge 
density has a finite continuum limit which is consistent with analytic predictions. 
Our study shows explicitly that some classically important features of an action are 
irrelevant for reaching the correct quantum continuum limit.

\end{abstract}

\newpage
 
\section{Introduction}

Universality is a key concept in classical statistical mechanics and in quantum 
field theory. In particular, in lattice field theory numerous lattice actions
yield the same universal continuum limit. It is well known that locality is vital for the viability of universality. A universality class is characterized
by the space-time dimension and the symmetries of the relevant fields. In order 
to construct lattice theories that fall into a desired universality class, one 
often imposes additional features of the corresponding classical theory on the 
lattice action. Usually one constructs a lattice action by replacing 
derivatives of the continuum fields by finite differences of the lattice 
fields. Such a discretization ensures the correct classical continuum limit. In 
addition, the lattice theory can then be investigated using perturbation 
theory. For example, for QCD it has been proved rigorously that the lattice 
regularization yields the same continuum limit as perturbative regularization 
schemes, such as dimensional regularization \cite{Rei88,Rei89}. 
Taking advantage of universality, and following Symanzik's improvement program 
\cite{Sym83,Sym83a}, one can systematically construct improved lattice actions 
\cite{Lue85,Lue85a} which eliminate lattice artifacts up to a given order in 
the lattice spacing. At a fixed point of the renormalization group, so-called 
classically perfect lattice actions have been constructed, which are free of 
lattice artifacts at the classical level \cite{Has94,Has02,Bur96,Bur01}. In 
particular, in asymptotically free theories, including QCD and the 2-d $O(3)$ 
model, a classically perfect fixed point action has been constructed by solving 
a minimization problem. In this paper, we proceed in a very different 
direction. In order to test the robustness of universality, we explicitly 
violate classically important properties of the action, such as the classical 
continuum limit, the applicability of perturbation theory, or the Schwarz 
inequality between the action and the topological charge. Hence, in contrast to 
Symanizik's lattice actions, which may, for example, be 1-loop improved, the 
actions that we will study can be viewed as tree-level impaired, but they are certainly still local. As we will 
see, even without appropriate classical features, the lattice theory acquires 
the correct quantum continuum limit. This also holds in one dimension ({\it 
i.e.}\ in quantum mechanics), although in this case one does usually not rely 
on universality. From the point of view of Wilson's renormalization group applied to critical phenomena the irrelevance of classical features of a given lattice action is perhaps not too surprising. For example, it is well known that the 4-d Ising model, whose classical Hamiltonian does not have a meaningful continuum limit, is in the same universality class as the $\phi^4$ quantum field theory. 

We will investigate local lattice actions that are invariant against small continuous
deformations of the lattice fields. Such actions --- which we will call 
{\em topological lattice actions} --- have infinitely many flat directions 
because they do not suppress small field fluctuations. As a consequence, they 
do not have the correct classical continuum limit and perturbation theory is 
not applicable. Depending on the nature of a topological action, it may or 
may not obey a Schwarz inequality. In this paper, we study two different types 
of topological lattice actions. The first one constrains the angle $\varphi$ 
between nearest-neighbor $O(N)$ spins to $|\varphi| < \delta$. All field 
configurations that satisfy this constraint have the same action value $S = 0$. 
Besides not having the correct classical continuum limit, this lattice action 
violates the Schwarz inequality between action and topological charge. The 
quantum continuum limit is reached by sending the maximally allowed angle 
$\delta$ to zero. Patrascioiu and Seiler \cite{Pat92} as well as Aizenman 
\cite{Aiz94} have used an action with an angle constraint to simplify the proof 
of the existence of a massless phase in the 2-d $O(2)$ model. Furthermore, 
Patrascioiu and Seiler have also used an angle-constraint action in their 
search for a massless phase in the 2-d $O(3)$ model \cite{Pat93,Pat02}, while 
Hasenbusch used the same action to argue that $\RP(N-1)$ models are in the same 
universality class as $O(N)$ models \cite{Has96}. Refs.\ 
\cite{Pat93,Pat02,Has96} presented numerical evidence that the action with the 
angle constraint falls in the same universality class as the standard action. 
Our study will confirm these results and will extend them by studying the 
cut-off effects of this topological action, as well as by investigating the 
topological susceptibility and the correlator of the topological charge 
density. Lattice actions with a similar constraint have also been used in
\cite{Lue82,Bie95,Her99,Lue99,Lue00,Fuk03,Fuk04,Fuk06,Jan06}, however, not with 
an emphasis on the topological properties of some of these actions. The second 
topological lattice action that we consider receives local contributions from 
the absolute value of the topological charge density. This action does not have 
the correct classical continuum limit either, but it obeys a lattice Schwarz 
inequality. Irrespective of this, as we will see, the correct quantum continuum 
limit is reached for both topological actions. 

It should be pointed out that lattice theories with topological actions are not 
regularizations of the topological field theories that arise in the context of 
string theory or conformal field theory \cite{Wit88}. While those theories 
realize new universality classes, the theories with topological lattice actions 
studied here fall into standard universality classes, despite the fact that 
they violate basic principles of classical physics.

The $O(N)$ model in $(N-1)$ dimensions has a non-trivial topological charge
$Q \in \Pi_{N-1}[S^{N-1}] = \Z$. We will investigate the 1-d $O(2)$ and
the 2-d $O(3)$ model which have topological charges in $\Pi_1[S^1]$ and 
$\Pi_2[S^2]$, respectively. While the 1-d $O(2)$ model will be studied 
analytically, the 2-d $O(3)$ model is investigated using Monte Carlo 
simulations. The 2-d $O(3)$ model can also be viewed as the $N = 2$ member of 
the 2-d $\CP(N-1)$ model family \cite{DAd78}. The $\CP(N-1)$ manifold is the 
coset space $SU(N)/U(N-1) = S^{2N-1}/S^1$. Since $\Pi_2[S^{2N-1}] = \{0\}$, one 
obtains $\Pi_2[\CP(N-1)] = \Pi_1[S^1] = \Z$, {\it i.e.}\ all 2-d $\CP(N-1)$ 
models 
possess a non-trivial topological charge. Since they are asymptotically free, 
have an anomalously broken classical scale invariance and a dynamically 
generated mass gap, as well as instantons and $\theta$-vacuum states, 2-d 
$\CP(N-1)$ models share many features with 4-d Yang-Mills theories. This has
motivated their detailed study beyond perturbation theory. The $\theta$-vacuum physics of 4-d $SU(N)$ gauge theories and of 2-d $\CP(N-1)$ models has been reviewed in \cite{Vic09}.

Topological aspects of lattice $\CP(N-1)$ models have been investigated in 
\cite{Ber81,Lue82b,Sch82,Mue82,Lue83,Bie95,Bla96,Bur96,Vic99,Bur01,Azc07}. Depending on the lattice 
action and the lattice definition of the topological charge, the quantum 
continuum limit of the topological susceptibility 
$\chi_t = \langle Q^2 \rangle/V$, where $V$ is the space-time volume, may be
spoiled by short-distance lattice artifacts. These so-called dislocations have
topological charge $|Q| = 1$ and a minimal value of the lattice action. 
Semi-classical arguments, which are, however, not rigorous, suggest that 
$\chi_t$ may have a power-law divergence in the quantum continuum limit 
\cite{Lue82b}. This is expected to happen in the $\CP(2)$ model, when one 
uses the standard action in combination with the geometric definition of the 
topological charge \cite{Lue83}. This problem does not arise for $\CP(N-1)$ 
models with $N \geq 4$. Even in the $\CP(2)$ case, dislocations can be 
eliminated and the proper continuum limit of $\chi_t$ can be attained if one 
uses a modified lattice action \cite{Lue83}. Dislocations have also been 
identified in 4-d lattice Yang-Mills theory \cite{Lue82a,Pug89,Goe89}. Again, 
semi-classical arguments suggest that they may spoil the quantum continuum 
limit in $SU(2)$ and $SU(3)$ Yang-Mills theory, if the standard Wilson action 
is used in combination with the geometric definition of the topological charge 
\cite{Lue82,Lue82a,Phi86,Goe86,Goe87}. As in the $\CP(2)$ model, dislocations 
can be eliminated in $SU(2)$ and $SU(3)$ Yang-Mills theory by using an improved 
lattice action \cite{Goe89}. The situation is more subtle in the 2-d $O(3)$ (or 
equivalently $\CP(1)$) model. In this case, a semi-classical calculation in the 
continuum already yields a divergent topological susceptibility $\chi_t$ even 
in a small space-time volume \cite{Sch82}. This divergence is not caused by 
lattice artifacts, but is an intrinsic feature of the theory in the continuum 
limit. Hence, one concludes that a meaningful quantum continuum limit of 
$\chi_t$ does not exist in the 2-d $O(3)$ model. This is supported by a 
calculation of $\chi_t$ using a classically perfect lattice action in 
combination with a classically perfect topological charge \cite{Bla96}, which 
eliminates dislocations and indeed shows no power-like divergence of the 
topological susceptibility. However, $\chi_t$ still diverges logarithmically, 
and thus a meaningful quantum continuum limit is not reached for this quantity. 
As we will see, the logarithmic divergence of $\chi_t$ even arises for 
topological actions, although in that case one might have expected a power-law 
divergence due to dislocations. On the other hand, the correlator of the topological
charge density will turn out to have a finite continuum limit. The concepts of 
classical and even quantum perfect definitions of the topological charge have 
also been investigated analytically in the 1-d $O(2)$ model \cite{Bie97}. 

In QCD, Ginsparg-Wilson lattice quarks \cite{Gin82} obey 
an Atiyah-Singer index theorem even at finite lattice spacing \cite{Has98}.
Based on Ginsparg-Wilson lattice fermions, unambiguous definitions of the 
topological susceptibility which are free of short-distance singularities have 
been provided \cite{Giu04,Lue04}. They have been used in a derivation of the 
Witten-Veneziano formula \cite{Wit79,Ven79a,Ven79b} for the $\eta'$-meson mass 
in a fully regularized non-perturbative framework \cite{Giu02,Deb05,Lue10}.

This paper is organized as follows. Section 2 contains an
analytic investigation of the 1-d $O(2)$ model using two different topological
lattice actions: one that suppresses topological charges and one that does not. 
In both cases, the correct quantum continuum limit is obtained. In Section 3 we 
study the 1-d $O(3)$ model in a similar manner. Section 4 presents a Monte Carlo
study of the 2-d $O(3)$ model. Again, we use two different topological lattice 
actions: one that does and one that does not obey a Schwarz inequality. As 
before, the correct quantum continuum limit is reached in both cases. The 
Monte Carlo data for the topological susceptibility are consistent with a 
logarithmic divergence, while the correlator of the topological charge density
has a finite continuum limit. We summarize our results and draw conclusions
in Section 5.

\section{The 1-d $O(2)$ Model}

In this section we consider the 1-d $O(2)$ model as an analytically solvable
test case with two different topological lattice actions: one that explicitly
suppresses topological charges and one that does not. Remarkably, irrespective 
of this, and although the lattice theories do not yield the correct classical 
limit, they do have the correct quantum continuum limit.

\subsection{The 1-d $O(2)$ Model in the Continuum}

In this subsection, we analytically solve the 1-d $O(2)$ model in the 
continuum. The results will then be compared with those of the corresponding 
lattice models. The 1-d $O(2)$ model is equivalent to a quantum mechanical 
rotor. Let us consider a particle of mass $M$ on a circle of radius $R$, and 
thus with the moment of inertia $I = M R^2$. The Hamilton operator takes the 
form
\begin{equation}
H(\theta) = 
- \frac{1}{2 I} \left(\partial_\varphi - i \frac{\theta}{2 \pi}\right)^2,
\end{equation}
where $\varphi$ is the angle describing the position of the particle, and
$\theta$ is analogous to the vacuum angle in QCD. At finite temperature 
$T = 1/\beta$, the corresponding Euclidean continuum action is given by
\begin{equation}
S[\varphi] = \int_0^\beta dt \ \frac{I}{2} \dot \varphi^2 - i \theta Q[\varphi],
\end{equation}
where the topological charge takes the form
\begin{equation}
Q[\varphi] = \frac{1}{2 \pi} \int_0^\beta dt \ \dot \varphi \in \Pi_1[S^1] 
= \Z.
\end{equation}
The energy eigenfunctions of the Hamilton operator $H(\theta)$ are given by
\begin{equation}
\langle \varphi|m\rangle = \frac{1}{\sqrt{2 \pi}} \exp(i m \varphi),
\end{equation}
where $m \in \Z$ specifies the angular momentum, and the corresponding energy 
eigenvalues are
\begin{equation}
\label{spectrum}
E_m(\theta) = \frac{1}{2 I} \left(m - \frac{\theta}{2 \pi}\right)^2.
\end{equation}
The canonical partition function takes the form
\begin{equation}
Z(\theta) = \mbox{Tr} \exp(- \beta H(\theta)) = 
\sum_{m \in \Z} \exp(- \beta E_m(\theta)).
\end{equation}
The (not yet normalized) distribution of the topological charge $Q$ 
is obtained as a Fourier transform of $Z(\theta)$
\begin{equation}
p(Q) = \frac{1}{2 \pi} \int_{-\pi}^\pi d\theta \ Z(\theta) 
\exp(- i \theta Q) = \sqrt{\frac{2 \pi I}{\beta}} 
\exp\left(- \frac{2 \pi^2 I}{\beta} Q^2\right),
\end{equation}
and the topological susceptibility (evaluated at $\theta = 0$) reads
\begin{equation}
\chi_t = \frac{\langle Q^2 \rangle}{\beta} = \frac{1}{\beta}
\frac{\sum_{Q \in \Z} Q^2 p(Q)}{\sum_{Q \in \Z} p(Q)}.
\end{equation}
In the zero temperature limit $\beta \rightarrow \infty$ this expression 
reduces to
\begin{equation}
\label{chitcont}
\chi_t = \frac{1}{4 \pi^2 I}.
\end{equation}
The correlation length $\xi$ (again evaluated at $\theta = 0$) is determined by 
the gap between the ground state and the first excited state
\begin{equation}
\xi = \frac{1}{E_1(0) - E_0(0)} = 2 I,
\end{equation}
such that at zero temperature
\begin{equation}
\chi_t \xi = \frac{1}{2 \pi^2}.
\end{equation}
Indeed, in the 1-d $O(2)$ model the topological susceptibility is a quantity 
with a meaningful finite quantum continuum limit, which scales like the inverse
correlation length. As we will discuss later, this is not the case in the 2-d 
$O(3)$ model. 

It is interesting to minimize the action (at $\theta = 0$) in a given 
topological charge sector. In the 1-d $O(2)$ model the minimizing 
configurations take the form
\begin{equation}
\varphi(t) = \varphi(0) + \frac{2 \pi Q t}{\beta},
\end{equation}
and they have the action $2 \pi^2 I Q^2/\beta$. Consequently, the action and 
the topological charge of all configurations obey the inequality
\begin{equation}
\label{O2ineq}
S[\varphi] \geq \frac{2 \pi^2 I Q[\varphi]^2}{\beta}.
\end{equation}
It should be noted that, unlike instantons in 4-d non-Abelian gauge theories 
and in 2-d $\CP(N-1)$ models, in the 1-d $O(2)$ model the topologically 
non-trivial minimal action configurations are not concentrated at an instant in 
Euclidean time, but are homogeneously distributed over time. In this sense, 
they do not deserve to be called instantons.
 
\subsection{A Topological Lattice Action without Topological Charge Suppression}

Let us now consider a 1-d lattice $O(2)$ model with spin variables 
$\varphi_t \in ]-\pi,\pi]$, {\it i.e.}\ an XY model, with zero action for 
nearest neighbor spins with 
$|(\varphi_{t+a} - \varphi_t) \ \mbox{mod} \ 2 \pi| < \delta$ and infinite 
action for $|(\varphi_{t+a} - \varphi_t) \ \mbox{mod} \ 2 \pi| \geq \delta$. 
Here $a$ is the lattice spacing. The geometric definition of the topological 
charge is given by
\begin{equation}
Q[\varphi] = \frac{1}{2 \pi} \sum_t (\varphi_{t+a} - \varphi_t) \ \mbox{mod} \
2 \pi \in \Z,
\end{equation}
with $(\varphi_{t+a} - \varphi_t) \ \mbox{mod} \ 2 \pi \in  ]-\pi,\pi]$. The 
partition function takes the form
\begin{equation}
Z(\theta) = \mbox{Tr} \ T(\theta)^N,
\end{equation}
with $\beta = N a$. The transfer matrix $T(\theta)$ has the elements
\begin{equation}
\langle \varphi_t|T(\theta)|\varphi_{t+a} \rangle =
\exp\left(- i \frac{\theta}{2 \pi} (\varphi_{t+a} - \varphi_t) \
\mbox{mod} \ 2 \pi\right),
\end{equation}
for $|(\varphi_{t+a} - \varphi_t) \ \mbox{mod} \ 2 \pi| < \delta$. For 
$|(\varphi_{t+a} - \varphi_t) \ \mbox{mod} \ 2 \pi| \geq \delta$, on the other 
hand, the transfer matrix elements vanish.

The transfer matrix can be diagonalized by changing to a basis of angular 
momentum eigenstates
\begin{eqnarray}
\langle m|T(\theta)|m'\rangle&=&\frac{1}{(2 \pi)^2} \int_{-\pi}^\pi d\varphi_t
\int_{-\pi}^\pi d\varphi_{t+a} \langle m|\varphi_t\rangle
\langle \varphi_t|T(\theta)|\varphi_{t+a} \rangle \langle \varphi_{t+a}|m'\rangle
\nonumber \\
&=&\frac{1}{(2 \pi)^2} \int_{-\pi}^\pi d\varphi_t
\int_{\varphi_t - \delta}^{\varphi_t + \delta} d\varphi_{t+a} 
\exp\left(- i \frac{\theta}{2 \pi} (\varphi_{t+a} - \varphi_t) \ \mbox{mod} \
2 \pi \right) \nonumber \\ 
&\times&\exp\left(- i m \varphi_t + i m' \varphi_{t+a}\right) \nonumber \\
&=&\delta_{m m'} \ \frac{1}{2 \pi} \int_{- \delta}^{\delta} d\varphi 
\exp\left(- i \frac{\theta}{2 \pi} \varphi + i m \varphi\right). 
\end{eqnarray}
Hence, the eigenvalues of the transfer matrix are given by
\begin{eqnarray}
\exp(- a E_m(\theta))&=&\frac{1}{2 \pi} \int_{- \delta}^{\delta} d\varphi 
\exp\left(i \left(m - \frac{\theta}{2 \pi}\right)\varphi\right) =
\frac{\sin((m - \theta/2 \pi)\delta)}{(m - \theta/2 \pi)\pi} \nonumber \\
&=&\frac{\delta}{\pi}
\left[1 - \frac{\delta^2}{6} \left(m - \frac{\theta}{2 \pi}\right)^2 + 
{\cal O}(\delta^4)\right].
\end{eqnarray}
In the continuum limit, $a \rightarrow 0$, we thus obtain
\begin{equation}
E_m(\theta) - E_0(0) = 
\frac{\delta^2}{6 a} \left(m - \frac{\theta}{2 \pi}\right)^2 =
\frac{1}{2 I} \left(m - \frac{\theta}{2 \pi}\right)^2.
\end{equation}
Here we have identified the moment of inertia as
\begin{equation}
I = \frac{3 a}{\delta^2}.
\end{equation}
It should be noted that $\exp(- a E_0(0)) = \delta/\pi$, such that the ground
state energy diverges in the continuum limit. This is no problem because only 
energy differences are physically relevant (in this context). In order to reach 
finite results in the continuum limit $a \rightarrow 0$, we must put 
$\delta = \sqrt{3 a/I}$. Interestingly, in this limit, the
topological lattice model reproduces the continuum 1-d $O(2)$ model. However,
it should be noted that the lattice transfer matrix is not positive definite.
In particular, for $|m - \theta/2 \pi|\delta > \pi/2$ the transfer matrix
eigenvalue $\exp(- a E_m(\theta))$ becomes negative. This is not necessarily
problematical, as long as this behavior does not affect the continuum limit.
The transfer matrix eigenvalues which obey
\begin{equation}
\left|m - \frac{\theta}{2 \pi}\right| < \frac{\pi}{2 \delta} = \frac{\pi}{2} 
\sqrt{\frac{I}{3 a}},
\end{equation}
are positive. This condition is automatically satisfied in the
continuum limit $a \rightarrow 0$. We thus conclude that the topological model
indeed provides an adequate regularization of the 1-d $O(2)$ model.

It is interesting to investigate the cut-off effects of the topological lattice
action. It is well-known that for the standard action the lattice artifacts set
in at ${\cal O}(a^2)$. In particular, the dimensionless ratio of energy gaps of 
the first two excited states is given by
\begin{equation}
\frac{E^s_2(0) - E^s_0(0)}{E^s_1(0) - E^s_0(0)} = 
4 \left(1 - \frac{a^2}{{\xi_s}^2} - 3 \frac{a^3}{{\xi_s}^3} + \dots \right), 
\ \xi_s = \frac{1}{E_1^s(0) - E_0^s(0)},
\end{equation}
while for a classically perfect action the lattice artifacts are even 
exponentially suppressed \cite{Bie97} \footnote{In \cite{Bie97} the 
corresponding expressions look different because they are expressed in terms of
the continuum correlation length.}
\begin{equation}
\frac{E^c_2(0) - E^c_0(0)}{E^c_1(0) - E^c_0(0)} = 
4 \left(1 + \frac{4}{\pi} \sqrt{\frac{\xi_c}{\pi a}} 
\exp\left(- \frac{\pi^2 \xi_c}{4a}\right) + \dots\right), \
\xi_c = \frac{1}{E_1^c(0) - E_0^c(0)}.
\end{equation}
For the topological lattice action under consideration, the corresponding 
result takes the form
\begin{equation}
\frac{E_2(0) - E_0(0)}{E_1(0) - E_0(0)} = 
4 \left(1 + \frac{3 a}{5 \xi} + \dots \right).
\end{equation}
Because the topological lattice action does not obey the correct classical 
continuum limit, it suffers from strong lattice artifacts of ${\cal O}(a)$. It
should be noted that Symanzik's systematic effective theory for lattice
artifacts \cite{Sym83,Sym83a} is applicable in quantum field theory but not in 
quantum mechanics. Indeed, the terms of ${\cal O}(a)$ in the topological action 
and of ${\cal O}(a^3)$ in the standard action would be forbidden if Symanizik's 
theory would apply. The scaling behavior of the various lattice actions is 
illustrated in Figure 1.
\begin{figure}[htb]
\begin{center}
\includegraphics[width=0.7\textwidth,angle=270]{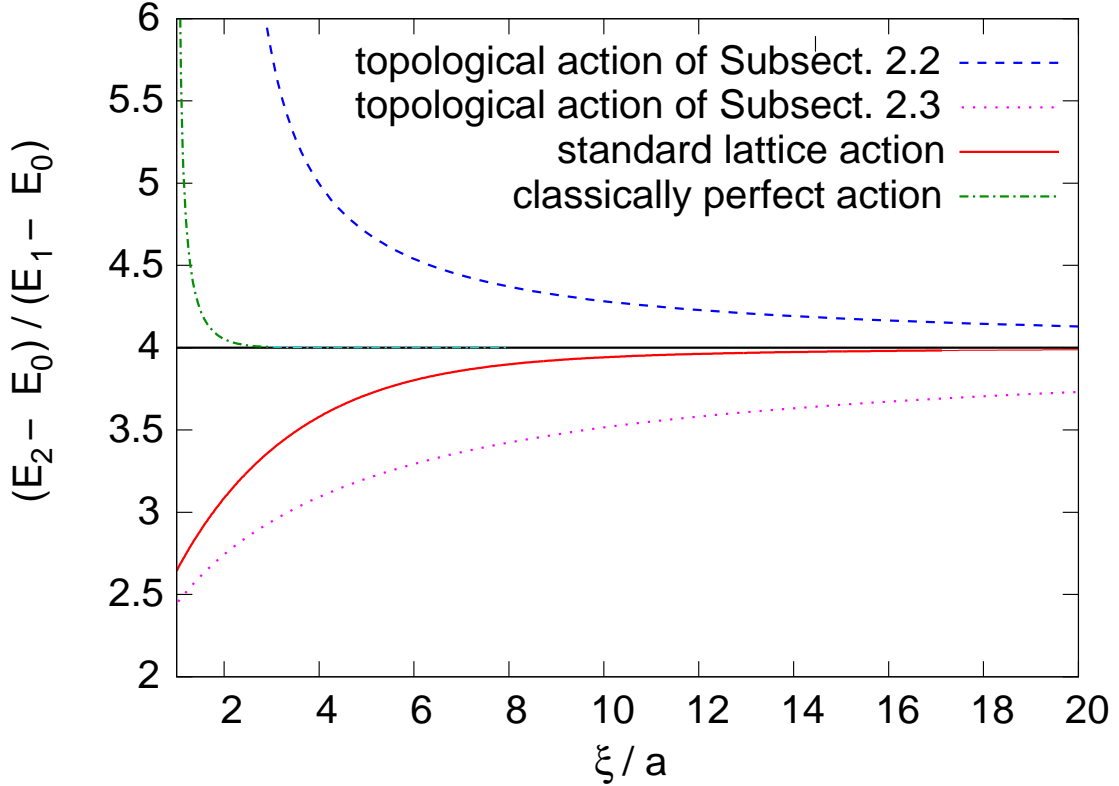}
\caption{\it The scaling behavior of the ratio $(E_2 - E_0)/(E_1 - E_0)$ of
energy gaps for different lattice actions: standard action (solid
curve), topological action with (dotted curve), and without topological
charge suppression (dashed curve), as well as classically perfect action 
(dashed-dotted curve). The topological lattice actions suffer from cut-off 
effects of ${\cal O}(a)$, while the standard action has only ${\cal O}(a^2)$ 
artifacts. For the classically perfect action, the lattice artifacts are even 
exponentially suppressed. In the continuum $(E_2 - E_0)/(E_1 - E_0) = 4$.} 
\end{center}
\end{figure}
Indeed, one sees that the results obtained with the topological lattice action
converge much slower than the ones for the standard or the classically perfect
action. Still, as pointed out before, the topological lattice action has the 
correct quantum continuum limit. 

Since the information about the topological charge distribution is encoded in
the $\theta$-dependence of the energy spectrum $E_m(\theta)$, the topological
susceptibility $\chi_t$ as well as other related topological quantities will
also automatically come out correctly. In order to show this explicitly, let us
also consider the partition function of the lattice model
\begin{equation}
Z(\theta) = \mbox{Tr} \ T(\theta)^N = \sum_{m \in \Z} \exp(- N a E_m(\theta)) =
\sum_{m \in \Z} 
\left[\frac{\sin((m - \theta/2 \pi)\delta)}{(m - \theta/2 \pi)\pi}\right]^N,
\end{equation}
such that
\begin{eqnarray}
p(Q)&=&\frac{1}{2 \pi} \int_{-\pi}^\pi d\theta \ Z(\theta) \exp(- i \theta Q)
\nonumber \\
&=&\frac{1}{2 \pi} \int_{-\infty}^\infty d\theta \ 
\left[\frac{2 \sin(\theta \delta/2 \pi)}{\theta}\right]^N \exp(- i \theta Q).
\end{eqnarray}
Since $p(Q)$ is the Fourier transform of an $N$-th power, it is given by the
$N$-fold convolution $p = p_1 \star p_1 \star \cdots \star p_1$. The elementary 
distribution is given by
\begin{equation}
p_1(Q) = \frac{1}{2 \pi} \int_{-\infty}^\infty d\theta \ 
\frac{2 \sin(\theta \delta/2 \pi)}{\theta} \exp(- i \theta Q) =
\Theta\left(Q + \frac{\delta}{2 \pi}\right) -
\Theta\left(Q - \frac{\delta}{2 \pi}\right). 
\end{equation}
In this case, $Q \in \R$ is not restricted to integer values, because the 
elementary contributions to the total topological charge originate from local
regions with open boundary conditions. We define the step function as 
$\Theta(Q) = 1$ for $Q > 0$, and as $\Theta(Q) = 0$ otherwise. In the 
zero-temperature limit $\beta \rightarrow \infty$, the topological 
susceptibility now takes the form
\begin{equation}
\chi_t = \frac{\int_{-\infty}^\infty dQ \ p_1(Q) Q^2}
{a \int_{-\infty}^\infty dQ \ p_1(Q)} = 
\frac{\int_{-\delta/2 \pi}^{\delta/2 \pi} dQ \ Q^2}
{a \int_{-\delta/2 \pi}^{\delta/2 \pi} dQ} = 
\frac{1}{4 \pi^2} \frac{\delta^2}{3 a} = \frac{1}{4 \pi^2 I},
\end{equation}
which is indeed the correct result of eq.(\ref{chitcont}) for the 1-d $O(2)$ 
model in the continuum.

It is interesting to note that this lattice action violates the inequality
(\ref{O2ineq}). This is obvious, because all allowed configurations have 
zero action. Still, for a given value of $\delta$ and for a given inverse
temperature $\beta = N a$, the allowed topological charges are restricted to
\begin{equation}
|Q[\varphi]| \leq \frac{\delta}{2 \pi} N = \frac{\delta \beta}{2 \pi a},
\end{equation}
such that
\begin{equation}
\frac{2 \pi^2 I Q[\varphi]^2}{\beta} \leq \frac{\pi I \delta^2 \beta}{2 a^2} =
\frac{3 \pi \beta}{2 a} \rightarrow \infty.
\end{equation}
Hence, unlike in the continuum theory, topologically non-trivial field
configurations are not suppressed by the lattice action in the continuum limit
$a \rightarrow 0$. Remarkably, nevertheless the lattice theory has the correct 
quantum continuum limit.

As a further scaling test, we compare the cut-off effects of the topological 
susceptibility in units of the mass gap, {\it i.e.}\ $\chi_t \xi$, for the 
various lattice actions. For the standard action one obtains
\begin{equation}
\chi_t^s \ \xi_s = \frac{1}{2 \pi^2} \left(1 + \frac{a^2}{3 {\xi_s}^2} + \frac{a^3}{{\xi_s}^3}+ 
\dots \right).
\end{equation}
while for the classically perfect action one finds
\begin{equation}
\chi_t^c \ \xi_c = \frac{1}{2 \pi^2}\left(1 - \sqrt{\frac{\pi \xi_c}{a}}
\left[1 - \frac{4}{\pi^2}\right] 
\exp\left(- \frac{\pi^2 \xi_c}{4 a}\right) + \dots \right).
\end{equation}
For the topological action, on the other hand, we obtain
\begin{equation}
\chi_t \xi = \frac{1}{4 \pi^2} \frac{\delta^2}{3} 
\left[\log\frac{\delta}{\sin\delta}\right]^{-1} = 
\frac{1}{2 \pi^2} \left(1 - \frac{a}{5 \xi} + \dots \right).
\end{equation}
As before the lattice artifacts of the topological action are of ${\cal O}(a)$,
while they are of ${\cal O}(a^2)$ for the standard action and exponentially
suppressed for the classically perfect action. The results for the various
actions are illustrated in Figure 2.
\begin{figure}[htb]
\begin{center}
\includegraphics[width=0.7\textwidth,angle=270]{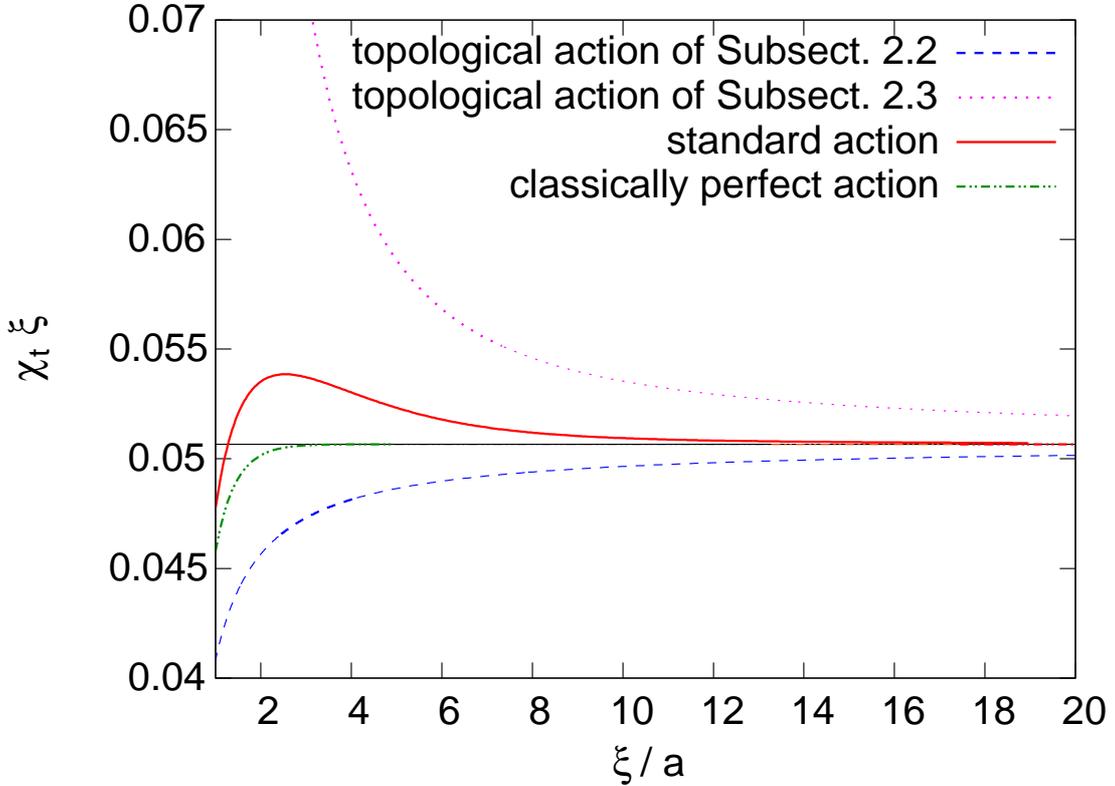}
\caption{\it The scaling behavior of the product $\chi_t \xi$ of the 
topological susceptibility and the correlation length for different lattice 
actions: standard action (solid curve), topological action with (dotted 
curve) and without topological charge suppression (dashed curve), as well 
as classically perfect action (dashed-dotted curve). The topological lattice 
actions suffer from cut-off effects of ${\cal O}(a)$, while the standard action 
has only ${\cal O}(a^2)$ artifacts. For the classically perfect action, the 
lattice artifacts are exponentially suppressed. The continuum value 
amounts to $\chi_t \xi = 1/(2 \pi^2)$.} 
\end{center}
\end{figure}

\subsection{A Topological Lattice Action with Topological Charge Suppression}

Let us now consider a topological lattice action which receives local
contributions from the topological charge density, {\it i.e.}
\begin{equation}
S[\varphi] = \lambda \sum_t |(\varphi_{t+a} - \varphi_t) \ \mbox{mod} \ 2 \pi|. 
\end{equation}
A (dimensionless) coupling constant $\lambda > 0$ suppresses configurations 
with non-zero topological charge. In particular, by construction this action 
obeys the inequality
\begin{equation}
\label{SQineq}
S[\varphi] \geq 2 \pi \lambda |Q[\varphi]|,
\end{equation}
which is inconsistent with the corresponding inequality (\ref{O2ineq}) of the 
continuum theory.

In this case, the transfer matrix takes the form
\begin{equation}
\langle \varphi_t|T(\theta)|\varphi_{t+a} \rangle = 
\exp(- \lambda |(\varphi_{t+a} - \varphi_t) \ \mbox{mod} \ 2 \pi|) 
\exp\left(- i \frac{\theta}{2 \pi} (\varphi_{t+a} - \varphi_t) \
\mbox{mod} \ 2 \pi\right).
\end{equation}
It is again diagonalized in the basis of angular momentum eigenstates and the
corresponding eigenvalues are given by
\begin{eqnarray}
\exp(- a E_m(\theta))&=&\frac{1}{2 \pi} \int_{- \pi}^\pi d\varphi 
\exp(- \lambda |\varphi|)
\exp\left(i \left(m - \frac{\theta}{2 \pi}\right)\varphi\right) \nonumber \\
&=&\frac{1}{\pi} \frac{1}{\lambda^2 + (m - \theta/2 \pi)^2} 
\Big\{\lambda - \exp(- \pi \lambda) \nonumber \\
&\times&\left[\lambda \cos\left(m \pi - \frac{\theta}{2}\right)
- \left(m - \frac{\theta}{2 \pi}\right) 
\sin\left(m \pi - \frac{\theta}{2}\right)\right]\Big\}.
\end{eqnarray}
For large $\lambda$, {\it i.e.}\ in the continuum limit, one then obtains
\begin{equation}
E_m(\theta) - E_0(0) = \frac{1}{a} 
\log\left[1 + \frac{1}{\lambda^2}\left(m - \frac{\theta}{2 \pi}\right)^2\right]
\ \rightarrow \ \frac{1}{a \lambda^2}\left(m - \frac{\theta}{2 \pi}\right)^2.
\end{equation}
In order to match the continuum result of eq.(\ref{spectrum}), we thus identify
\begin{equation}
I = \frac{a \lambda^2}{2} \ \Rightarrow \ \lambda = \sqrt{\frac{2 I}{a}}.
\end{equation}
Using the inequality (\ref{SQineq}), one then concludes that the action of
topologically non-trivial field configurations diverges in the continuum limit,
\begin{equation}
S[\varphi] \geq 2 \pi \lambda |Q[\varphi]| = 
2 \pi \sqrt{\frac{2 I}{a}} |Q[\varphi]| \ \rightarrow \ \infty.
\end{equation}
Remarkably, despite this fact, in the quantum
continuum limit the $\theta$-dependent energy spectrum still agrees with the 
one of the continuum theory.

As before, we consider the lattice artifacts of the ratio of energy gaps, which 
now takes the form
\begin{equation}
\frac{E_2(0) - E_0(0)}{E_1(0) - E_0(0)} = 
4 \left(1 - \frac{3 a}{2 \xi} + \dots\right).
\end{equation}
Again, the lattice artifacts, which are also illustrated in Figure 1, are of 
${\cal O}(a)$. The artifacts of the topological lattice action with topological
charge suppression are even a factor $5/2$ larger than for the topological 
action without topological charge suppression. In particular, even for 
$\xi = 15 a$ the deviations from the continuum limit are as large as 10 percent,
while they are only about 0.5 percent with the standard action.

Let us again consider the $\theta$-dependent partition function
\begin{eqnarray}
Z(\theta)&=&\mbox{Tr} \ T(\theta)^N = \sum_{m \in \Z} \exp(- N a E_m(\theta)) =
\nonumber \\
&=&\sum_{m \in \Z} \left\{\frac{1}{\pi} \frac{1}{\lambda^2 + 
(m - \theta/2 \pi)^2} \{\lambda - \exp(- \pi \lambda) \right. \nonumber \\
&\times&\left.\left[\lambda \cos\left(m \pi - \frac{\theta}{2}\right)
- \left(m - \frac{\theta}{2 \pi}\right) 
\sin\left(m \pi - \frac{\theta}{2}\right)\right]\}\right\}^N.
\end{eqnarray}
As before, the corresponding topological charge distribution $p(Q)$ is obtained
as the $N$-fold convolution of an elementary distribution
\begin{eqnarray}
p_1(Q)&=&\int_{-\infty}^\infty d\theta \ \frac{2}{(2 \pi \lambda)^2 + \theta^2} 
\left\{\lambda - \exp(- \pi \lambda) \left[\lambda \cos\frac{\theta}{2} -
\frac{\theta}{\pi} \sin\frac{\theta}{2}\right]\right\} \exp(- i \theta Q) 
\nonumber \\
&=&[1 - \exp(- 2 \pi \lambda)] \exp(- 2 \pi \lambda |Q|).
\end{eqnarray}
The topological susceptibility in the zero-temperature limit is then given by
\begin{equation}
\chi_t = \frac{\int_{-\infty}^\infty dQ \ p_1(Q) Q^2}
{a \int_{-\infty}^\infty dQ \ p_1(Q)} = \frac{1}{2 \pi^2 \lambda^2 a} =
\frac{1}{4 \pi^2 I},
\end{equation}
which again is the correct quantum continuum limit.

Let us again consider the lattice artifacts of the product $\chi_t \xi$. Up to
exponentially small corrections, we obtain
\begin{equation}
\chi_t \xi = \frac{1}{2 \pi^2 \lambda^2} 
\left[\log\left(1 + \frac{1}{\lambda^2}\right)\right]^{-1} = 
\frac{1}{2 \pi^2} \left(1 + \frac{a}{2 \xi} + \dots \right).
\end{equation}
As for the topological action of the previous subsection, the lattice artifacts
are of ${\cal O}(a)$. This result is also illustrated in Figure 2.

We hence conclude that, despite the fact that the two topological lattice 
actions do not have the correct classical continuum limit, cannot
be treated perturbatively, or violate the classical inequality 
(\ref{O2ineq}), they both have the correct quantum continuum limit. In
particular, this holds for the $\theta$-dependent energy spectrum and all
quantities derived from it, including the topological susceptibility.

\section{The 1-d $O(3)$ Model}

In this section we consider an angle-constraint topological lattice action for 
the 1-d $O(3)$ model, which describes a quantum mechanical particle moving on 
the surface of a sphere $S^2$. Again, despite the fact that the topological 
lattice action does not obey the correct classical continuum limit, it 
correctly reproduces the quantum continuum limit.

\subsection{A Particle Moving on $S^2$}

Let us now consider a particle of mass $M$ moving on a sphere $S^2$ of radius 
$R$. The Hamiltonian then takes the form
\begin{equation}
H = \frac{{\vec L}^2}{2 I},
\end{equation}
where $\vec L$ is the angular momentum and $I = M R^2$ is the moment of inertia.
The corresponding eigenfunctions are the spherical harmonics 
$\langle \theta,\varphi|l m\rangle = Y_{lm}(\theta,\varphi)$ with
$l \in \{0,1,2,\dots\}$ and $m \in \{-l,-l+1,\dots,l\}$, and the $(2l+1)$-fold
degenerate energy eigenvalues are
\begin{equation}
\label{O3spectrum}
E_l = \frac{l(l+1)}{2 I}.
\end{equation}
In this case, one cannot construct a topological charge and the Euclidean
action is simply given by
\begin{equation}
S[\vec e] = \int dt \ \frac{I}{2} \partial_t \vec e \cdot \partial_t \vec e,
\end{equation}
with $\vec e(t) = (\sin\theta(t) \cos\varphi(t),\sin\theta(t) \sin\varphi(t),
\cos\theta(t))$.

\subsection{The 1-d $O(3)$ Model with a Topological Lattice Action}

Let us now consider the $O(3)$ model with spins $\vec e_t =
(\sin\theta_t \cos\varphi_t,\sin\theta_t \sin\varphi_t,\cos\theta_t)$
attached to the
sites $t$ of a 1-d lattice with spacing $a$. The lattice action constrains the
angle between neighboring spins $\vec e_t$ and $\vec e_{t+a}$ to a maximal value
$\delta$, {\it i.e.}\ $\vec e_t \cdot \vec e_{t+a} \geq \cos\delta$. The action 
vanishes
as long as this constraint is satisfied and is infinite otherwise. The 
corresponding transfer matrix is then given by
$\langle \theta_t,\varphi_t|T|\theta_{t+a},\varphi_{t+a}\rangle = 1$ for
$\vec e_t \cdot \vec e_{t+a} \geq \cos\delta$, and 
$\langle \theta_t,\varphi_t|T|\theta_{t+a},\varphi_{t+a}\rangle = 0$
otherwise. We now put $\vec e_t = (0,0,1)$ and $\vec e_{t+a} = 
(\sin\theta \cos\varphi,\sin\theta \sin\varphi,\cos\theta)$. Inserting
complete sets of states $|l m\rangle$ and using the fact that the transfer 
matrix is $O(3)$-invariant, {\it i.e.}
\begin{equation}
\langle l m|T|l' m'\rangle = \delta_{ll'} \delta_{mm'} \exp(- a E_l),
\end{equation}
one obtains
\begin{eqnarray}
\Theta(\cos\theta - \cos\delta)&=&\langle 0,0|T|\theta,\varphi\rangle =
\sum_{l,m} \langle 0,0|l m\rangle \langle l m|T|l m\rangle 
\langle l m|\theta,\varphi\rangle \nonumber \\
&=&\sum_{l,m} Y_{l m}(0,0) \exp(- a E_l) Y_{l m}(\theta,\varphi)^*.
\end{eqnarray}
Inverting this relation, we find
\begin{eqnarray}
Y_{l m}(0,0) \exp(- a E_l)&=&\int_{-1}^1 d\cos\theta \int_0^{2\pi} d\varphi \
\Theta(\cos\theta - \cos\delta) Y_{l m}(\theta,\varphi) \nonumber \\
&=&2 \pi \delta_{m,0} \int_{\cos\delta}^1 dx \ \sqrt{\frac{2 l + 1}{4 \pi}}
P_l(x),
\end{eqnarray}
where $P_l(x)$ is a Legendre polynomial. Using 
$Y_{l m}(0,0) = \delta_{m,0} \sqrt{(2 l + 1)/4 \pi}$ and applying the Legendre
differential equation (with $P_l'(x) = dP_l(x)/dx$)
\begin{equation}
\frac{d}{dx} \left[(1 - x^2) P_l'(x)\right] + l(l+1) P_l(x) = 0,
\end{equation}
for $l \neq 0$ one obtains
\begin{eqnarray}
\exp(- a E_l)&=&2 \pi \int_{\cos\delta}^1 dx \ P_l(x) = - \frac{2\pi}{l(l+1)}
(1 - x^2) P_l'(x)\vert_{\cos\delta}^1 \nonumber \\
&=&\frac{2\pi}{l(l+1)} \sin^2\delta \ P_l'(\cos\delta).
\end{eqnarray}
Similarly, for $l=0$ one finds $\exp(- a E_0) = 2 \pi (1 - \cos\delta)$,
which results in
\begin{equation}
E_l - E_0 = - \frac{1}{a} \log\left[\frac{1 + \cos\delta}{l(l+1)} \
P_l'(\cos\delta)\right].
\end{equation}
Expanding in small values of $\delta$, {\it i.e.}\ putting $\cos\delta \approx
1 - \delta^2/2$, one obtains
\begin{eqnarray}
E_l - E_0&=&- \frac{1}{a} \log\left[\frac{2 - \delta^2/2}{l(l+1)} \
P_l'(1 - \delta^2/2)\right] \nonumber \\
&=&- \frac{1}{a} \log\left\{\frac{2}{l(l+1)} \left[P_l'(1) - 
\frac{\delta^2}{4} \left[P_l'(1) + 2 P_l''(1)\right]
\right]\right\}  + {\cal O}(\delta^4) \nonumber \\
&=&\frac{\delta^2}{8a} l(l+1) + {\cal O}(\delta^4).
\end{eqnarray}
Here we have used
\begin{equation}
P_l'(1) = \frac{l(l+1)}{2}, \ 
P_l'(1) + 2 P_l''(1) = \left(\frac{l(l+1)}{2}\right)^2. 
\end{equation}
Hence, identifying
\begin{equation}
\label{coupling}
I = \frac{4 a}{\delta^2},
\end{equation}
we indeed reproduce the correct spectrum of eq.(\ref{O3spectrum}) in the 
quantum continuum limit. It turns out that in the 1-d $O(N)$ model the moment of
inertia is given by $I = (N+1)a/\delta^2$.

Let us again consider the lattice artifacts in the ratio of energy gaps,
which now takes the form
\begin{equation}
\frac{E_2(0) - E_0(0)}{E_1(0) - E_0(0)} = 
3 \left(1 + \frac{a}{3 \xi} + \dots \right).
\end{equation}
As for the 1-d $O(2)$ model, the lattice artifacts are of ${\cal O}(a)$.

\section{The 2-d $O(3)$ Model}

In this section we consider the 2-d $O(3)$ model using numerical simulations. 
We first summarize the results obtained before in the continuum and with 
standard as well as modified lattice actions. In analogy to Subsections
2.2 and 2.3 we then investigate two topological lattice actions, one that does 
and one that does not obey a Schwarz inequality between action and topological 
charge. Remarkably, in both cases we will again obtain the correct quantum 
continuum limit. While it is well-known that in this model the topological 
susceptibility is logarithmically divergent, we will see that the correlator of 
the topological charge density has a finite continuum limit. Hence there are topological quantities that do have a finite continuum limit in this model.

\subsection{The 2-d $O(3)$ Model in the Continuum and with the Standard Lattice 
Action}

In this subsection we summarize results obtained before either in the continuum
or using the standard lattice action for the 2-d $O(3)$ model. In the 
continuum, the 2-d $O(3)$ model has the Euclidean action
\begin{equation}
S[\vec e] = \frac{1}{2 g^2} \int d^2x \ 
\partial_\mu \vec e \cdot \partial_\mu \vec e - i \theta Q[\vec e].
\end{equation}
Here $\vec e(x) \in S^2$ is a 3-component unit-vector field defined at each
point $x$ in a 2-dimensional Euclidean space-time, and the topological charge
is given by
\begin{equation}
Q[\vec e] = \frac{1}{8 \pi} \int d^2x \ \varepsilon_{\mu\nu} 
\vec e \cdot (\partial_\mu \vec e \times \partial_\nu \vec e) \in \Pi_2[S^2] 
= \Z.
\end{equation}
It is straightforward to show that the following integral, which is 
non-negative by construction, takes the form
\begin{equation}
I = \int d^2x \ (\partial_\mu \vec e \pm 
\varepsilon_{\mu\nu} \vec e \times \partial_\nu \vec e)^2 =
4 g^2 S[\vec e] \pm 16 \pi Q[\vec e] \geq 0. 
\end{equation}
This immediately implies the Schwarz inequality
\begin{equation}
\label{Schwarz}
S[\vec e] \geq \frac{4 \pi}{g^2} |Q[\vec e]|.
\end{equation}
Field configurations which saturate this inequality are (anti-)self-dual, 
{\it i.e.}
\begin{equation}
\partial_\mu \vec e = \pm 
\varepsilon_{\mu\nu} \vec e \times \partial_\nu \vec e,
\end{equation}
and are known as (anti-)instantons. For these configurations the Lagrangian
${\cal L}(\vec e)$ (at $\theta = 0$) is proportional to the absolute value of 
the topological charge density $q(\vec e)$, {\it i.e.}
\begin{equation}
\label{eqsq}
{\cal L}(\vec e) = 
\frac{1}{2 g^2} \partial_\mu \vec e \cdot \partial_\mu \vec e =
\frac{1}{2 g^2} \big\vert \varepsilon_{\mu\nu} 
\vec e \cdot (\partial_\mu \vec e \times \partial_\nu \vec e) \big\vert =
\frac{4 \pi}{g^2} |q(\vec e)|.
\end{equation}
In Section 4.3 we will introduce a topological lattice action such that 
${\cal L}(\vec e)$ is proportional to $|q(\vec e)|$ for all configurations, not just 
for instantons or anti-instantons.

Remarkably, at $\theta = 0$ the 2-d $O(3)$ model can be solved exactly using the
Bethe ansatz \cite{Pol83,Wie85,Zam79}. Based on these results, using the 
Wiener-Hopf technique, the exact mass gap of the 2-d $O(3)$ model 
\begin{equation}
m = \frac{8}{e} \Lambda_{\overline{MS}},
\end{equation}
has been derived \cite{Has90}. Here $e$ is the base of the
natural logarithm, and $\Lambda_{\overline{MS}}$ is the scale generated by
dimensional transmutation in the modified minimal subtraction 
renormalization scheme. Even the finite-size effects of the mass gap $m(L)$
for the 2-d $O(3)$ model on a finite periodic spatial interval of size $L$
have been calculated analytically \cite{Bal04}. A dimensionless physical
quantity 
\begin{equation}
u_0 = L m(L) = L/\xi(L),
\end{equation}
is then obtained as the ratio of the spatial size $L$ and the 
finite-volume correlation length $\xi(L) = 1/m(L)$. Similarly, one defines
the step scaling function (with scale factor 2)
\begin{equation}
\sigma(2,u_0) = 2 L m(2 L),
\end{equation}
which is thus also known analytically. Later, we will compare Monte Carlo 
data for $\sigma(2,u_0)$, obtained with a topological lattice action, with
the analytic result. The step scaling function was first studied on 
the lattice by L\"uscher, Weisz, and Wolff \cite{Lue91} using the 
standard action
\begin{equation}
S[\vec e] = \frac{1}{g^2} \sum_{x,\mu} \vec e_x \cdot \vec e_{x+\hat\mu}.
\end{equation}
Now $x$ denotes the lattice sites, and $\hat\mu$ is a vector of length $a$
pointing in the $\mu$-direction. The step scaling function is affected by 
lattice artifacts which --- for some time --- seemed not to be described by 
Symanzik's effective theory. Interestingly, a recent careful study of the 
lattice artifacts has shown that large logarithms arise, and that Symanzik's 
theory does describe the lattice artifacts correctly \cite{Bal09,Bal10}. Hence, 
one may conclude that the continuum limit of the finite-volume mass gap $m(L)$ 
is finally well understood. Furthermore, excellent agreement between the 
Zamolodchikov bootstrap S-matrix and Monte Carlo data, obtained with both the 
standard and a classically perfect lattice action, has been reported 
\cite{Bal99}.

Let us now discuss the topological susceptibility 
\begin{equation}
\chi_t = \frac{\langle Q^2 \rangle}{V}
\end{equation}
of the 2-d $O(3)$ model. Here $V = \beta L$ is the space-time volume. Based on 
naive dimensional analysis, one would expect that $\chi_t \xi^2$ approaches a 
constant in the continuum limit. While this expectation is met in 2-d $\CP(N-1)$
models with $N \geq 3$, the $\CP(1)$ (or equivalently $O(3)$) model behaves 
differently. The pathological behavior of $\chi_t$ in the 2-d $O(3)$ model
manifests itself already in the continuum formulation. As was pointed out by
L\"uscher \cite{Lue82b}, the integration over the instanton size parameter
$\rho$ in a semi-classical calculation gives rise to a logarithmically 
divergent ultra-violet contribution to $\chi_t$ which is proportional to $\int d\rho/\rho$. 

Berg and L\"uscher have investigated $\chi_t$ using a geometric definition of
the lattice topological charge \cite{Ber81}. In this definition, each plaquette 
of the square lattice is divided into two triangles, as illustrated in Figure 
3. 
\begin{figure}[htb]
\begin{center}
\includegraphics[width=0.5\textwidth,angle=270]{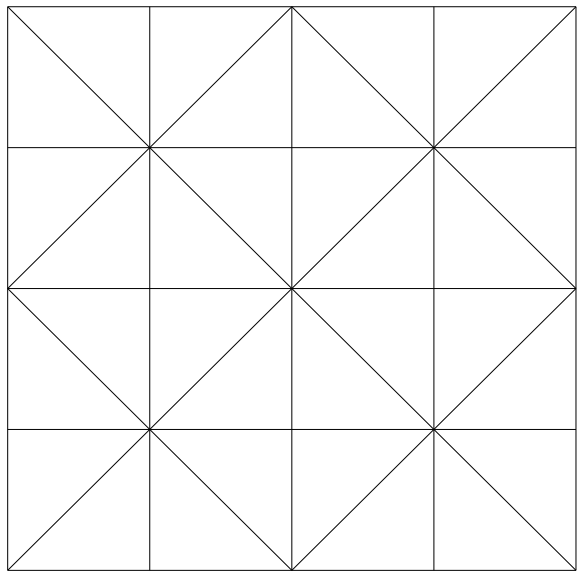}
\caption{\it Decomposition of the square lattice into triangles to be used to define the action and the topological charge.}
\end{center}
\end{figure}
The spins $\vec e_1$, $\vec e_2$, and $\vec e_3$ at the three corners of a 
lattice triangle $t_{123}$ define the corners of a spherical triangle on $S^2$. 
The oriented area $A_{123}$ of the spherical triangle is given by
\begin{eqnarray}
\label{area}
&&A_{123} = 2 \varphi \in [- 2 \pi,2 \pi], \quad 
X + i Y = r \exp(i \varphi), \nonumber \\
&&X = 1 + 
\vec e_1 \cdot \vec e_2 + \vec e_2 \cdot \vec e_3 + \vec e_3 \cdot \vec e_1,
\quad Y = \vec e_1 \cdot (\vec e_2 \times \vec e_3).
\end{eqnarray}
The geometric topological charge is the sum of the oriented areas $A_{xyz}$ 
over all triangles $t_{xyz}$, normalized by the area $4 \pi$ of $S^2$, 
{\it i.e.}
\begin{equation}
Q[\vec e] = \frac{1}{4 \pi} \sum_{t_{xyz}} A_{xyz} \in \Z.
\end{equation}
The decomposition of the square lattice into triangles illustrated in Figure 3 
is invariant under $\pi/2$ rotations and translation invariant by an even 
number of lattice spacings. Obviously, the topological charge inherits these 
symmetries.
\begin{figure}[htb]
\begin{center}
\includegraphics[width=0.45\textwidth]{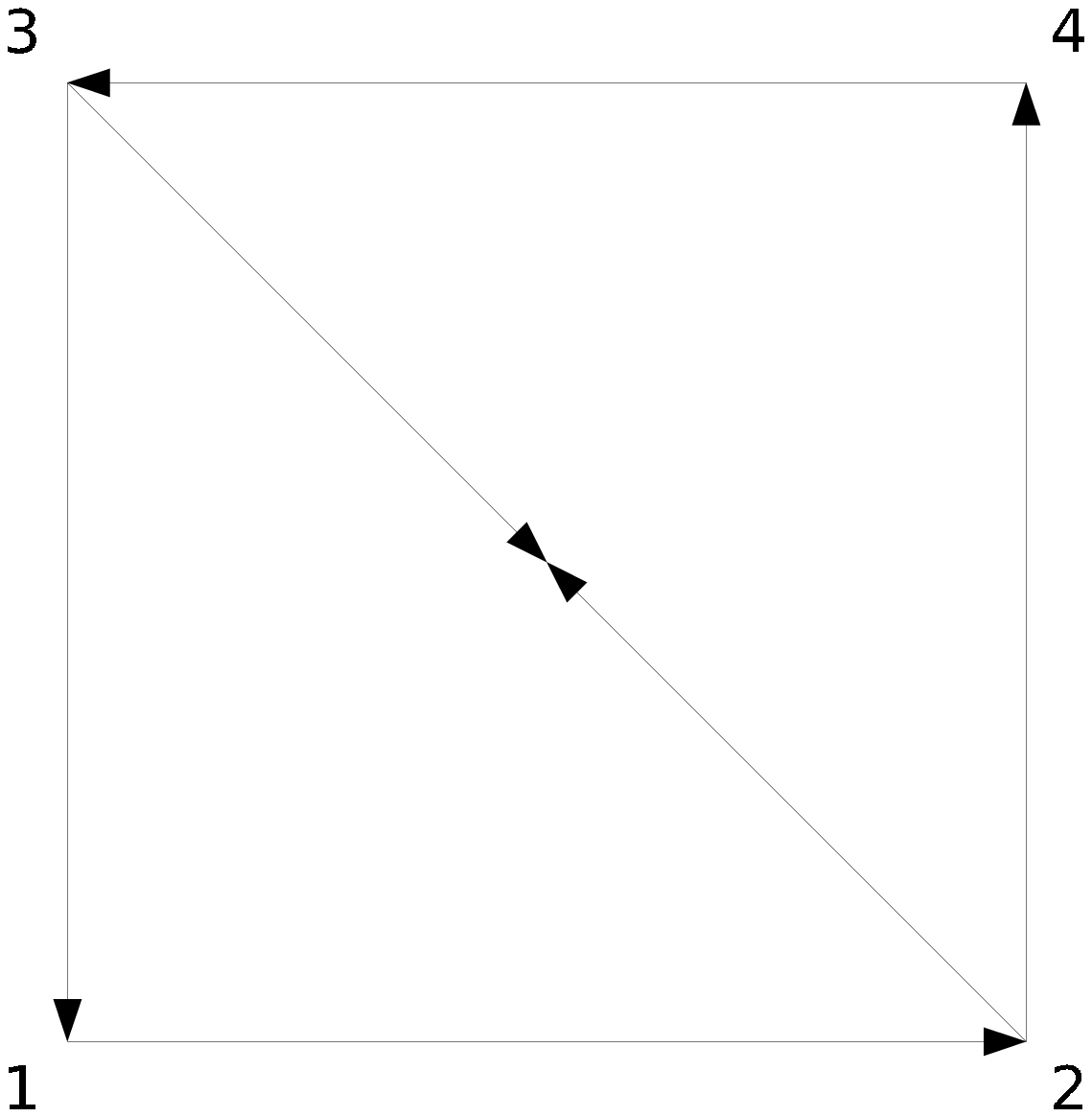} \hskip0.5cm
\includegraphics[width=0.45\textwidth]{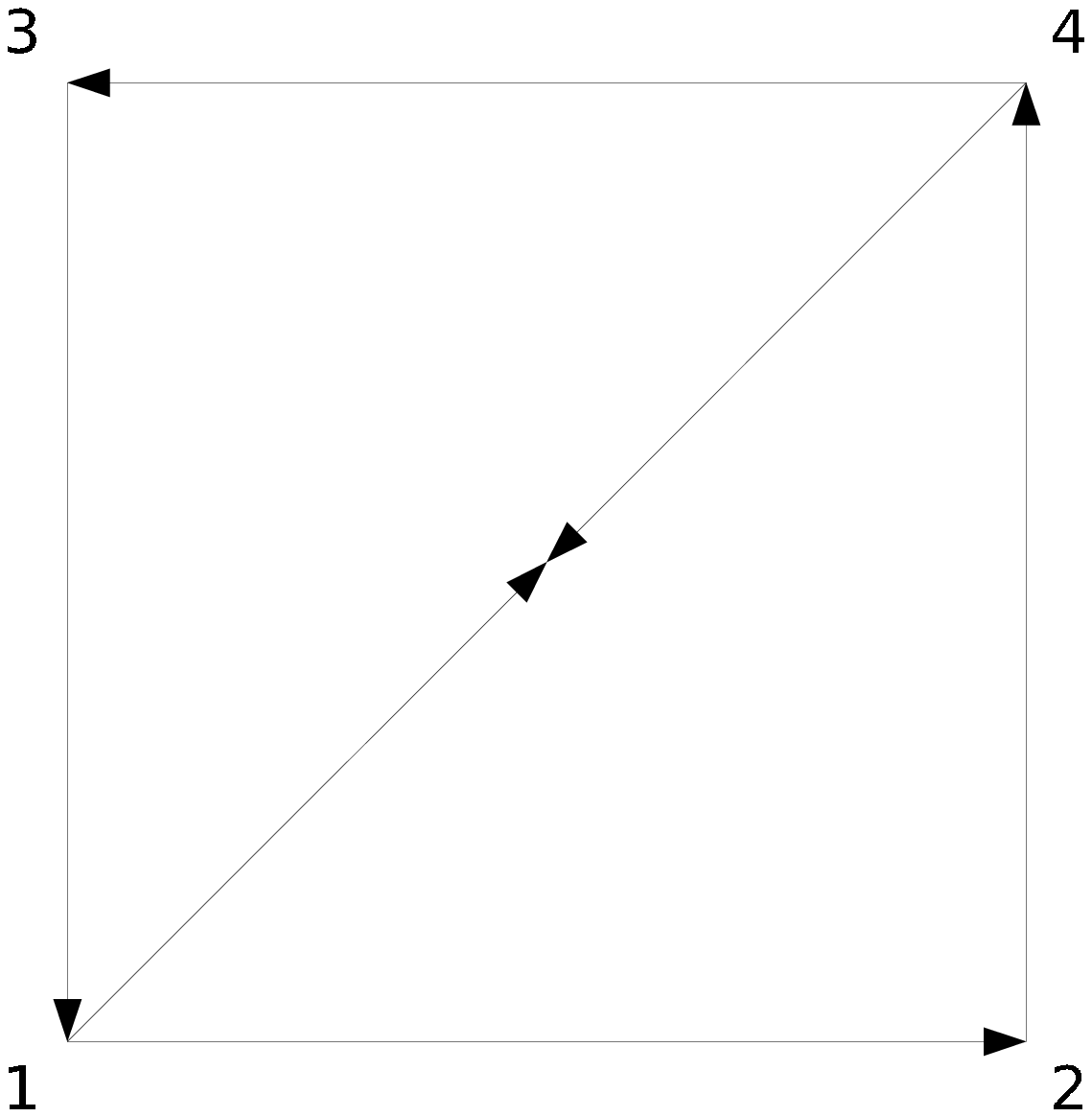}
\caption{\it A given plaquette 1234 can be decomposed into two triangles in two
alternative ways: a) 123 and 243 or b) 124 and 143.}
\end{center}
\end{figure}
As illustrated in Figure 4, an individual lattice plaquette 
1234 can be divided into two triangles 123 and 243 or alternatively into
triangles 124 and 143. In general, it is not guaranteed that $A_{123} + A_{243}$ 
is equal to $A_{124} + A_{143}$. However, if the relative angle between 
nearest-neighbor spins is smaller than $\pi/2$, one can indeed show that 
$A_{123} + A_{243} = A_{124} + A_{143}$. In that case, the topological charge 
becomes independent of the particular decomposition into triangles, and thus 
becomes invariant even against translations by a single lattice spacing. Here
we prefer to work on a triangulated quadratic (rather than a triangular) lattice
because the definition of the step scaling function refers to a rectangular 
space-time volume.

The geometric topological charge is undefined for a set of exceptional field 
configurations, which form a set of measure zero in configuration space. These
exceptional configurations contain a spherical triangle that covers exactly one
half of $S^2$ and thus has an area $\pm 2 \pi$. The infinitesimal neighborhood
of an exceptional configuration contains configurations whose topological 
charges differ by 1. If the relative angle between nearest-neighbor spins is 
smaller than $\pi/2$, exceptional configurations cannot arise. When one uses 
the standard lattice action, exceptional configurations have a finite action.
In the following subsection, we will consider a topological lattice action that
constrains the relative angle between nearest-neighbor spins to a maximum value
$\delta$. When $\delta < \pi/2$, exceptional configurations are excluded, and 
--- just as in the continuum --- different topological sectors are then 
separated by infinite-action barriers. This means that the angle-constraint
topological action, but not the standard action, naturally leads to a unique 
segmentation of configuration space into distinct topological sectors.

In addition to the logarithmic ultra-violet divergence that is present in the 
continuum theory, in the 2-d lattice $O(3)$ model $\chi_t$ is affected by 
additional short-distance artifacts. These so-called dislocations are minimal
action field configurations with a non-zero topological charge. As such, they
depend on both the definition of the lattice action and the definition of the
lattice topological charge. Using the standard lattice action in combination 
with the geometric definition of the lattice topological charge, the
dislocations are exceptional configurations with an action $S_d = c/g^2$ with 
$c \approx 6.69 < 4 \pi$ \cite{Lue82b}. Semi-classical arguments (which, 
however, are not rigorous) suggest that the topological susceptibility should 
scale as $\chi_t \propto \exp(- S_d)$. Due to asymptotic freedom, the
correlation length scales as $\xi/a \propto \exp(2 \pi/g^2)$, where $2 \pi$ is
the universal 1-loop coefficient of the $\beta$-function. Hence, one expects
a power-law divergence of the dimensionless combination
\begin{equation}
\chi_t \xi^2 \propto \exp(- S_d) \exp\left(\frac{4 \pi}{g^2}\right) =
\exp\left(\frac{4 \pi - c}{g^2}\right) \propto 
\left(\frac{\xi}{a}\right)^{2 - c/2 \pi}.
\end{equation}
Dislocations have been eliminated in \cite{Bla96} by using a classically 
perfect lattice action in combination with a classically perfect definition of 
the lattice topological charge. Perfect discretizations are based on the 
renormalization group and eliminate cut-off effects at the classical level. 
In particular, one then has $S_d = 4 \pi/g^2$, {\it i.e.}\ $c = 4 \pi$. 
Interestingly, the topological susceptibility was then found to still diverge 
logarithmically. From all this one concludes that $\chi_t$ does not have a 
finite continuum limit in the 2-d $O(3)$ model. 

The situation is different in 2-d $\CP(N-1)$ models with $N \geq 3$. While 
$\chi_t$ is still divergent in the $\CP(2)$ model when one uses the standard
lattice action in combination with the geometric definition of the topological
charge, dislocations can be suppressed by choosing a modified lattice action
\cite{Lue82b,Lue83}. For $N \geq 4$, dislocations do not cause any problems, 
even when the standard action is used. Similarly, in 4-d $SU(N)$ lattice 
Yang-Mills theories, when one uses the geometric topological charge, $\chi_t$ 
suffers from dislocations for $N = 2$ and 3, which again can be suppressed by 
using a modified lattice action \cite{Pug89,Goe89}.

\subsection{A Topological Lattice Action without Topological Charge Suppression}

In analogy to the 1-d $O(2)$ and $O(3)$ models, we now investigate the 2-d 
$O(3)$ model with a topological lattice action that constrains the relative 
angle of nearest-neighbor spins on a square lattice to a maximum angle 
$\delta$. All configurations that violate this constraint are forbidden and 
thus have 
infinite action, while all other configurations are allowed and have zero 
action. This action can be simulated with the very efficient Wolff cluster 
algorithm \cite{Wol89,Wol90}. Two neighboring spins are put in the same cluster 
if flipping one of them on a randomly chosen reflection plane would increase 
their relative angle beyond $\delta$. For the efficiency of the algorithm it 
is essential that, using this method, only spins on the same side of the 
reflection plane end up in the same cluster. The correlation function of two 
spins $\langle \vec e_x \cdot \vec e_y \rangle$ can then be computed very 
accurately using an improved estimator. 

We have fine-tuned the maximal angle $\delta$ such that the finite-volume mass 
gap $m(L)$ satisfies $u_0 = L m(L) = 1.0595$ for lattices with 
$L/a = 10, 16, 32$, and 64. The corresponding values of $\delta$ are listed in
Table 1. 
\begin{table}
\begin{center}
\begin{tabular}{|c|c|c|c|}
\hline
$L/a$ & $\delta/\pi$ & $\xi(L)/a$ & $\xi(2L)/a$ \\
\hline
\hline
10 & 0.44858728 & 9.4383(1)  &  15.9515(15) \\
\hline
16 & 0.434009   & 15.1013(2) &  25.4787(10) \\
\hline
32 & 0.415095   & 30.2029(2) &  50.847(3)   \\
\hline
64 & 0.398665   & 60.406(1)  & 101.578(4)   \\
\hline
\end{tabular}
\caption{\it Constraint angle $\delta$ that leads to $u_0 = L m(L) = 1.0595$
for various lattice sizes together with the correlation lengths $\xi(L)$ and
$\xi(2L)$.}
\end{center}
\end{table}
By measuring the mass gap $m(2L)$ for these values, we have 
then determined the lattice value $\Sigma(2,u_0,a/L) = 2L m(2L)$ of the step 
scaling function, which is known to approach the continuum limit 
$\sigma(2,u_0 = 1.0595) = 1.26121035(2)$ \cite{Bal09,Bal10}. 
\begin{figure}[htb]
\begin{center}
\includegraphics[width=0.77\textwidth,angle=270]{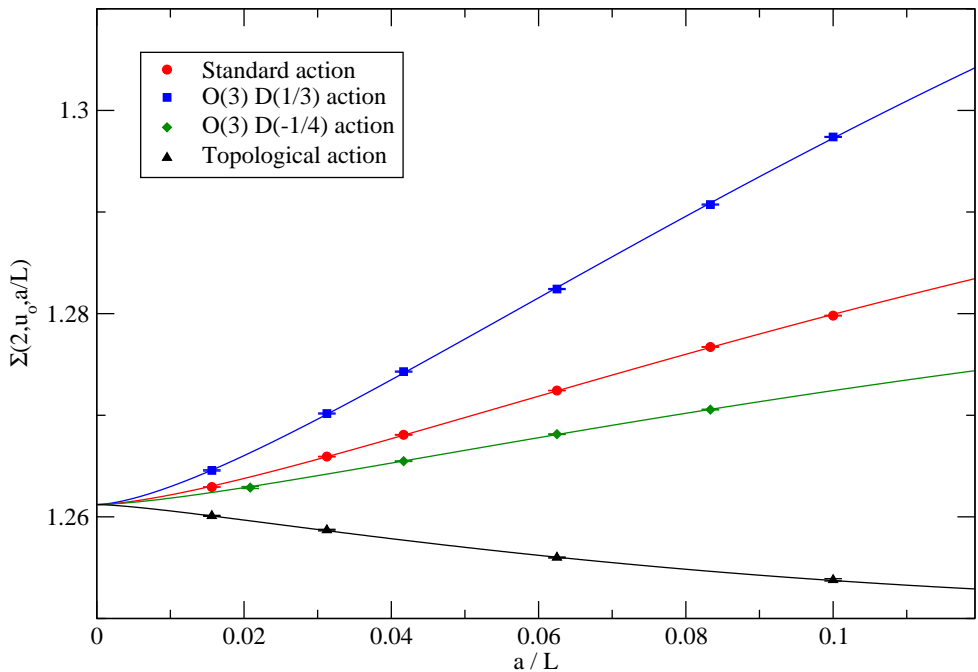}
\caption{\it Cut-off dependence of the step scaling function 
$\Sigma(2,u_0,a/L)$ at $u_0 = 1.0595$ for four different lattice actions: the 
standard as well as two different modified actions ($O(3) D(1/3)$ and 
$O(3) D(-1/4)$) \cite{Bal09}, and the topological lattice action without 
topological charge suppression. The lines are fits based on Symanzik's 
effective theory with the continuum value fixed to the exact result
$\Sigma(2,u_0 = 1.0595,a/L \rightarrow 0) = \sigma(2,u_0 = 1.0595) = 
1.26121035(2)$.} 
\end{center}
\end{figure}
Figure 5 compares the cut-off effects of $\Sigma(2,u_0,a/L)$ for the 
topological action with the standard and with two modified actions. At first 
glance, it seems that all four actions have lattice artifacts of 
${\cal O}(a)$. This would contradict Symanzik's effective theory. However, as 
investigated in detail \cite{Bal09,Bal10}, the standard and the modified 
actions indeed have lattice artifacts of ${\cal O}(a^2 \log^3(a/L))$, with 
large logarithmic corrections mimicking ${\cal O}(a)$ effects. In particular, 
Symanzik's theory correctly describes the observed lattice artifacts. Since 
lattice perturbation theory is not applicable to topological lattice actions, 
one might think that one cannot use Symanzik's theory to predict the lattice 
artifacts. However, this is true only to some extent. Symanzik's effective theory is 
formulated in the continuum and should be applicable to any lattice theory that 
reaches the correct quantum continuum limit. As we have seen analytically for 
the 1-d $O(2)$ and $O(3)$ models, the topological lattice actions suffer from 
lattice artifacts of ${\cal O}(a)$, which would contradict Symanzik's effective 
theory. However, since the underlying power-counting does not work in quantum 
mechanics, Symanzik's theory is not applicable in that case. In the 2-d case, 
however, Symanzik's theory applies even to topological lattice actions. This 
suggests to fit the lattice step scaling function to
\begin{equation}
\Sigma(2,u_0 = 1.0595,a/L) = \sigma(2,u_0 = 1.0595) + \frac{a^2}{L^2} 
\left[B \log^3(L/a) + C \log^2(L/a) + \dots\right],
\end{equation}
which gives a good fit for $B = - 0.067(4)$ and $C = 0.014(9)$. Interestingly, 
in the range considered here, the lattice artifacts of the topological action 
are smaller than those of the standard action, for which one obtains 
$B = 0.041(6)$ and $C = 0.26(2)$. In fact, for the standard action at $L/a = 64$
the sub-leading term proportional to $\log^2(L/a)$ is still larger than the 
leading term proportional to $\log^3(L/a)$, while this is not the case for the
topological action. One can estimate that the lattice artifacts of the standard
action will be smaller than the ones of the topological action only for 
correlation lengths larger than about $5 \times 10^4 a$. Furthermore, if one 
uses $\sigma(2,u_0 = 1.0595)$ as a fit parameter, only the topological action 
data are consistent with the exact value within error bars, while the other 
actions give rise to small deviations. Given the fact that the topological 
action violates the classical continuum limit, and is thus tree-level impaired,
it performs remarkably well. This may perhaps encourage the use of topological 
lattice actions also in other models, including Abelian and non-Abelian gauge 
theories.

The very accurate approach to the exact continuum result for 
$\sigma(2,u_0 = 1.0595)$ strongly suggests that the topological lattice action 
indeed leads into the standard universality class of the 2-d $O(3)$ model. This
confirms earlier results of \cite{Pat92,Pat93,Has96,Pat02} and also justifies a 
posteriori the use of a topological lattice action in \cite{Bie95}.

\subsection{A Topological Lattice Action with Topological Charge Suppression}

In analogy to the action for the 1-d $O(2)$ model discussed in Section 2.3, we 
now introduce a topological lattice action for the 2-d $O(3)$ model which 
explicitly suppresses topological charges and obeys a Schwarz inequality. 
Again, the action is given by the absolute value of the topological charge 
density, {\it i.e.}
\begin{equation}
\label{topact3}
S[\vec e] = \lambda \sum_{t_{xyz}} |A_{xyz}|.
\end{equation}
Here $|A_{xyz}|$ is the area of the spherical triangle on $S^2$ defined by the 
spins $\vec e_x$, $\vec e_y$, and $\vec e_z$ at the three corners of a lattice 
triangle $t_{xyz}$, {\it cf.} eq.(\ref{area}), and $\lambda$ is a positive coupling 
constant. By construction, this action obeys the inequality
\begin{equation}
S[\vec e] \geq \lambda \Big\vert\sum_{t_{xyz}} A_{xyz}\Big\vert = 
4 \pi \lambda |Q[\vec e]|. 
\end{equation}
A comparison with the Schwarz inequality eq.(\ref{Schwarz}) of the continuum
theory may suggest to identify $\lambda = 1/g^2$, but, as we will see, this is
not necessarily justified. By construction, for this
topological lattice action the Lagrangian ${\cal L}(\vec e)$ is proportional to
the absolute value of the topological charge density, {\it i.e.}\ 
${\cal L}(\vec e) = 4 \pi \lambda |q(\vec e)|$, for all configurations. In the 
continuum theory the corresponding eq.(\ref{eqsq}) is satisfied only for 
instantons or anti-instantons.
It is interesting to investigate the limit $\lambda \rightarrow \infty$. Then 
the allowed configurations only contain spherical triangles of zero area.
Consequently, at $\lambda = \infty$ all spins $\vec e_x$ fall on a common great
circle in $S^2$, and thus seem to represent an $O(2)$ model. When a triangle 
$t_{xyz}$ contains an $O(2)$ vortex, the corresponding area is 
$|A_{xyz}| = 2 \pi$. Hence, at $\lambda = \infty$, the 2-d $O(3)$ model from 
above reduces to a 2-d $O(2)$ model from which vortices have been eliminated. 
Such a model is expected to be in a massless phase. In the 2-d $O(3)$ model, 
the continuum limit is approached at $\lambda \rightarrow \infty$, not by 
putting $\lambda = \infty$. In particular, universality suggests that we should 
still recover the asymptotically free continuum limit of the 2-d $O(3)$ model.

Unfortunately, the action of eq.(\ref{topact3}) cannot be simulated with an
efficient Wolff-type embedding cluster algorithm. While it is possible to define
a cluster algorithm that is ergodic and obeys detailed balance, one is forced to
put spins in one common cluster although they are on different sides of the
reflection plane. This renders the algorithm inefficient. Hence we have used
a Metropolis algorithm to simulate this action. A high-precision study of the
mass gap, as described in the previous subsection, is then not feasible. A 
numerically better accessible quantity is the second moment correlation
length, which does not require a fit of the correlation function 
$G(x-y) = \langle \vec e_x \cdot \vec e_y \rangle$ at large distances. Instead 
one considers the Fourier transform
\begin{equation}
\widetilde G(p) = \sum_x G(x) \exp(i p x)
\end{equation}
for a quadratic $L \times L$ lattice, which yields the susceptibility 
$\chi = \widetilde G(p=0)$ as well as the corresponding quantity at the smallest
non-zero momentum $F = \widetilde G(p=(2\pi/L,0))$. The second moment 
correlation length is then defined as
\begin{equation}
\label{xi2}
\xi_2(L) = \left(\frac{\chi - F}{4 F \sin^2(\pi/L)}\right)^{1/2}.
\end{equation}
Using the corresponding mass $m_2(L) = 1/\xi_2(L)$ one can define the step 
scaling function for the second moment correlation length as
$\Sigma_2(2,m_2(L)L,a/L) = 2L m_2(2L)$. In the continuum limit, 
$\xi_2(L)/a \rightarrow \infty$, this function is again universal and has been 
determined in \cite{Car95}. The universality has been verified in the $(2+1)$-d 
spin $1/2$ quantum Heisenberg model \cite{Bea98}, which dimensionally reduces 
to the 2-d $O(3)$ model in the low-temperature limit \cite{Has91}. We have 
measured the step scaling function $\Sigma_2(2,m_2(L)L,a/L)$ on quadratic 
lattices with $L/a = 50$ and 100. In addition, we have measured this quantity 
for the topological lattice action of the previous subsection, which does not 
explicitly suppress topological charges. As illustrated in Figure 6, in both 
cases one finds very good agreement with the step scaling function obtained 
with the standard action. This again confirms that topological lattice actions 
lead to the correct quantum continuum limit, despite the fact that the 
classical continuum limit is not correctly represented.
\begin{figure}[htb]
\begin{center}
\includegraphics[width=0.78\textwidth,angle=270]{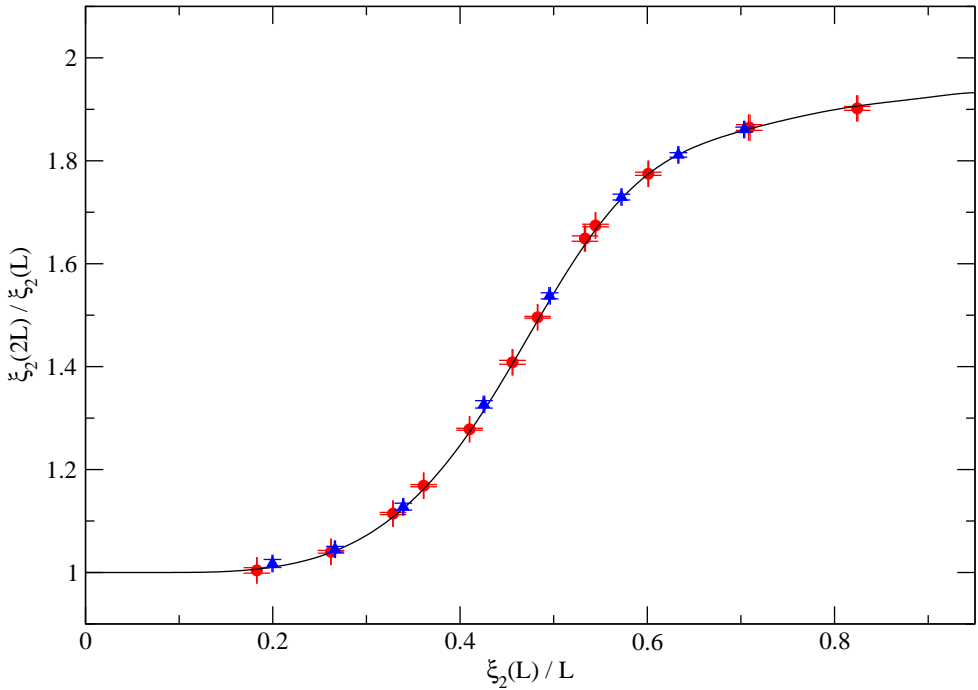}
\caption{\it The ratio of second moment correlation lengths
$\xi_2(2 L)/\xi_2(L)$ obtained with the topological lattice action with 
(triangles) and without topological charge suppression (circles). Within error
bars the data fall onto the universal curve that was extracted from simulations
with the standard action \cite{Car95}.}
\end{center}
\end{figure}

\subsection{Topological Susceptibility from Topological Lattice Actions}

As discussed in the introduction, in the 2-d $O(3)$ model the topological
susceptibility $\chi_t$ does not have a finite continuum limit. When the 
standard lattice action is used, based on (non-rigorous) semi-classical 
arguments one would expect a power-law divergence of $\chi_t$ due to 
dislocations which are short-range lattice artifacts carrying non-zero 
topological charge \cite{Lue82b}. Even when a classically perfect action is 
used in combination with a classically perfect topological charge, $\chi_t$ 
still diverges logarithmically \cite{Bla96}. The logarithmic divergence is not 
a lattice artifact, but is an inherent feature of the 2-d $O(3)$ model even in 
the continuum. In this subsection we investigate the topological susceptibility 
using the two topological lattice actions with and without explicit topological 
charge suppression.

First, we consider the topological lattice action of Subsection 4.2 which does
not explicitly suppress topological charges and which violates the Schwarz 
inequality. In order to hold the physical volume fixed while approaching the 
continuum limit, we demand $L = 4 \xi_2(L)$.  Since it is easier to compute 
numerically, we base this criterion on the second moment correlation length 
$\xi_2(L)$, and not on the inverse mass gap $\xi(L)$, which is consistently just
about 1.7 percent larger than $\xi_2(L)$. In the infinite volume limit the ratio $\xi/\xi_2 = 1.0007(1)$ has been determined very accurately in \cite{Cam97}. As illustrated in Figure 7, the discrepancy between this result and the ratio $\xi(L)/\xi_2(L) = 1.017(3)$ observed at $L = 4 \xi_2(L)$ can be attributed to finite volume effects. \footnote{We thank the referee for bringing this issue to our attention.}

\begin{figure}[htb]
\begin{center}
\includegraphics[width=0.7\textwidth,angle=270]{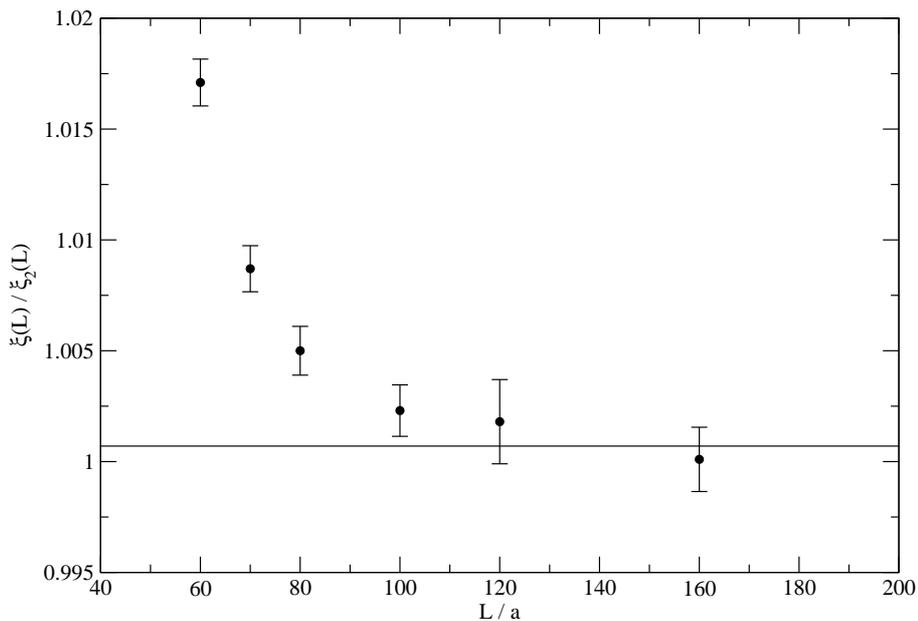}
\caption{\it The ratio $\xi(L)/\xi_2(L)$ as a function of $L/a$ (at fixed $\delta/\pi=0.4849$) is consistent with the infinite volume result $\xi/\xi_2 = 1.0007(1)$ obtained in \cite{Cam97}. The infinite volume result is represented by the horizontal line.}
\end{center}
\end{figure}

Figure 8 shows the finite volume 
correlation length $\xi_2(L)/a$ as a function of the constraint angle $\delta$,
together with a fit of the form
\begin{equation}
\label{xi2fit1}
\frac{\xi_2(L)}{a} = A g^2 \exp\left(\frac{2 \pi}{g^2}\right), \ 
\frac{1}{g^2} = \frac{b}{\delta^2} + c,
\end{equation}
which yields $A = 0.24(2)$, $b = 0.263(2)$, and $c = - 0.54(2)$, with 
$\chi^2/\mbox{d.o.f.} \approx 0.7$. Only the large lattices with $L/a \geq 200$ 
are included in the fit. The exponential increase of the correlation length is 
a consequence of asymptotic freedom. The fit assumes asymptotic scaling and 
contains the universal 1- and 2-loop coefficients of the $\beta$-function. The
relation between $\delta$ and $g$ is inspired by the corresponding relation
eq.(\ref{coupling}) in the 1-d $O(3)$ model. It should, however, be pointed out
that it is just a phenomenological ansatz which cannot be derived analytically,
because perturbation theory does not apply to topological lattice actions. For 
the standard action, asymptotic scaling is known to set in only at very large 
correlation lengths \cite{Car95}. Hence, also the value of $b$ determined here 
may not yet correspond to the asymptotic value in the continuum limit. Again, 
since perturbation theory is not applicable to topological lattice actions, one 
cannot determine the asymptotic value of $b$ analytically.  
\begin{figure}[htb]
\begin{center}
\includegraphics[width=0.7\textwidth,angle=270]{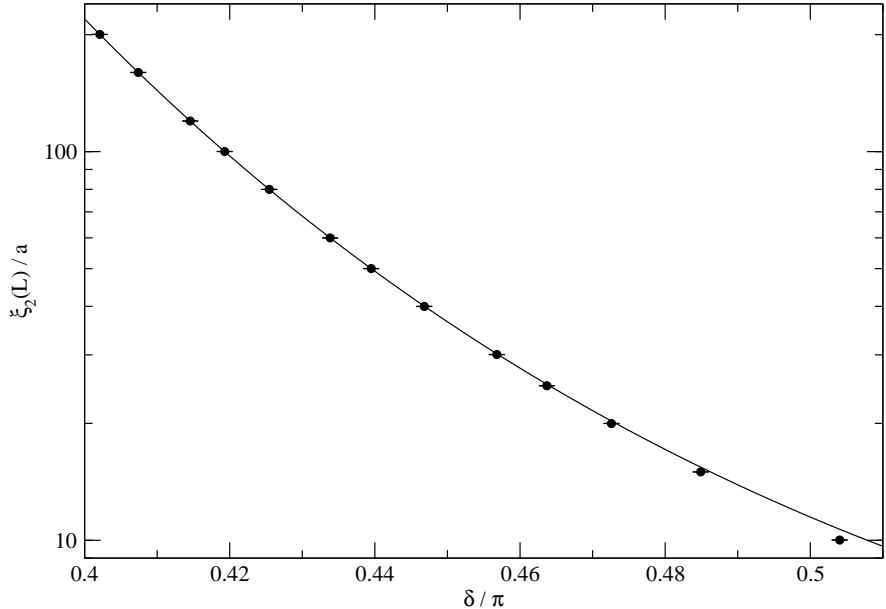}
\caption{\it Second moment correlation length $\xi_2(L)/a$ at fixed 
physical size $L = 4 \xi_2(L)$ as a function of the constraint angle 
$\delta$. The line is a fit of the form of eq.(\ref{xi2fit1}),
$\xi_2(L)/a =  A g^2 \exp(2 \pi/g^2)$ with $1/g^2 = b/\delta^2 + c$.}
\end{center}
\end{figure}

In order to investigate the scaling of the topological susceptibility, we 
consider the dimensionless combination
\begin{equation}
16 \chi_t(L) \xi_2(L)^2 = 16 \frac{\langle Q^2\rangle(L)}{L^2} 
\left(\frac{L}{4} \right)^2 = \langle Q^2\rangle(L),
\end{equation}
as we approach the continuum limit $\xi_2(L)/a \rightarrow \infty$. In Table 2 
we list the results of Monte Carlo simulations on various lattice sizes ranging 
from $L/a = 40$ to $L/a = 800$. 
\begin{table}
\begin{center}
\begin{tabular}{|c|c|c|c|c|}
\hline
$L/a$ & $\delta/\pi$ & $\xi_2(L)/a$ & $\xi(L)/a$ & $\langle Q^2 \rangle(L)$ \\
\hline
\hline
 40 & 0.50405 &  10.016(1) & 10.20(2) & 1.083(5) \\
\hline
 60 & 0.48490 &  14.996(8) & 15.29(4) & 1.299(6) \\
\hline
 80 & 0.47260 &  19.981(4) & 20.34(6) & 1.455(7) \\
\hline
100 & 0.46370 &  24.99(2)  & 25.46(8) & 1.598(9) \\
\hline
120 & 0.45680 &  30.061(4) & 30.57(8) & 1.72(2) \\
\hline
160 & 0.44680 &  39.98(2)  & 40.67(9) & 1.923(9) \\
\hline
200 & 0.43950 &  50.02(1)  & 50.75(9) & 2.09(2) \\
\hline
240 & 0.43385 &  59.98(2)  & 60.97(9) & 2.23(1) \\
\hline
320 & 0.42545 &  79.98(3)  & 81.18(9) & 2.45(2) \\
\hline
400 & 0.41930 & 100.07(3)  & 101.5(2) & 2.64(2) \\
\hline
480 & 0.41455 & 119.92(4)  & 121.8(3) & 2.78(2) \\
\hline
640 & 0.40740 & 159.87(7)  & 162.5(3) & 3.046(8)\\
\hline
800 & 0.40210 & 200.27(6)  & 203.7(3) & 3.22(2) \\
\hline
\end{tabular}
\caption{\it Monte Carlo data for the topological charge squared 
$\langle Q^2\rangle(L)$ at fixed physical size $L = 4 \xi_2(L)$ approaching 
the continuum limit $\xi_2(L)/a \rightarrow \infty$. Here $\xi_2(L)$ is the 
second moment correlation length, defined in eq.(\ref{xi2}), while $\xi(L)$ is 
the inverse mass gap, and $\delta$ is the constraint angle in the topological 
action of Subsection 4.2.}
\end{center}
\end{table}
Since this topological  action vanishes for all allowed configurations, the
dislocation action is $S_d = 0$. Hence, based on a naive semi-classical 
argument, one might expect $\chi_t \xi^2 \propto \exp(- S_d) (\xi/a)^2 = 
(\xi/a)^2$. Such a power-law divergence is not reflected by the numerical data 
depicted in Figure 9. Instead, the data are well fitted by
\begin{equation}
\label{Qfit}
\langle Q^2\rangle(L) = A \log\left(\frac{L + L_0}{L_1}\right),
\end{equation}
with $A = 0.91(1)$, $L_0/a = 39(2)$, $L_1/a = 24(1)$, and 
$\chi^2/\mbox{d.o.f.} \approx 0.5$. A power-law fit 
$\langle Q^2\rangle(L) = B L^\nu + C$ yields a small power $\nu = 0.21(2)$, and $\chi^2/\mbox{d.o.f.} \approx 1.9$.

\begin{figure}[htb]
\begin{center}
\includegraphics[width=0.7\textwidth,angle=270]{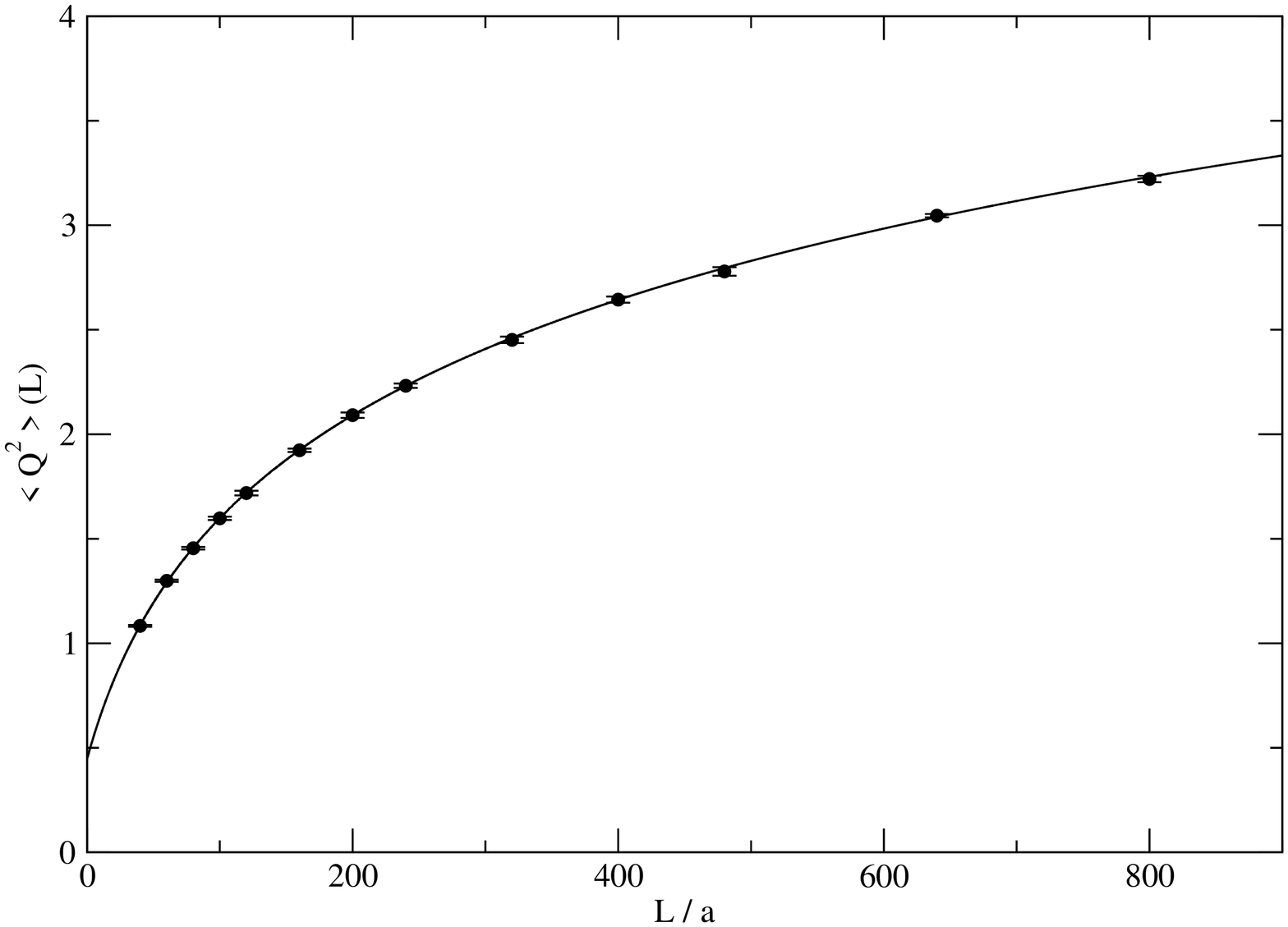}
\caption{\it The topological charge squared $\langle Q^2\rangle(L)$ at fixed 
physical size $L = 4 \xi_2(L)$, for the topological lattice action of
Subsection 4.2. As one approaches the continuum limit 
$\xi_2(L)/a \rightarrow \infty$, $\langle Q^2\rangle(L)$ diverges 
logarithmically. The line is a fit of the form of eq.(\ref{Qfit}), 
$\langle Q^2\rangle(L) = A \log((L + L_0)/L_1)$.}
\end{center}
\end{figure}

A logarithmic divergence of $\langle Q^2 \rangle(L)$ was already encountered in 
the continuum theory \cite{Sch82} and was also observed on the lattice using a 
classically perfect action \cite{Bla96}. It is interesting that the same 
behavior arises for the topological action which should be most vulnerable by 
dislocations. While the naive semi-classical argument is not rigorous, it is 
still remarkable that even the presence of zero-action dislocations does not 
spoil the logarithmic divergence of the continuum theory. This suggests that
dislocations may not cause power-law divergences in other cases, including 
2-d $\CP(N-1)$ models and 4-d non-Abelian gauge theories, either. Indeed, no 
such divergence has been detected, for example, in numerical data for $\chi_t$ 
in 4-d $SU(2)$ Yang-Mills theory \cite{Kro87}.

We have investigated $\langle Q^2 \rangle(L)$ at fixed physical size
$L = 4 \xi_2(L)$, also for the topological action of Subsection 4.3, which 
explicitly suppresses topological charges. The corresponding numerical results 
are listed in Table 3.
\begin{table}
\begin{center}
\begin{tabular}{|c|c|c|c|}
\hline
$L/a$ & $4 \pi \lambda$ & $\xi_2(L)/a$ & $\langle Q^2 \rangle(L)$ \\
\hline
\hline
 40 & 11.9215 &  10.00(2) & 1.102(2) \\
\hline
 60 & 13.781  & 14.98(2) & 1.282(2) \\
\hline
 80 & 15.112    & 20.00(4) & 1.419(3) \\
\hline
120 & 16.988  & 30.04(5) & 1.631(3) \\
\hline
160 & 18.325    & 40.00(6) & 1.784(4)\\
\hline
\end{tabular}
\caption{\it Monte Carlo data for the topological charge squared 
$\langle Q^2\rangle(L)$ at fixed physical size $L = 4 \xi_2(L)$ approaching 
the continuum limit $\xi_2(L)/a \rightarrow \infty$. Here $\xi_2(L)$ is the 
second moment correlation length and $\lambda$ is the coupling constant in the 
topological action of Subsection 4.3.}
\end{center}
\end{table}
Since in this case no efficient cluster algorithm is available, the calculation 
is limited to lattices up to $L/a = 160$. As illustrated in Figure 10, the data 
for $\langle Q^2 \rangle(L)$ are again consistent with the logarithmic
divergence of eq.(\ref{Qfit}). Here the best fit yields
$A = 0.63(2)$, $L_0/a = 20(3)$, $L_1/a = 10(1)$, with
$\chi^2/\mbox{d.o.f.} \approx 0.4$.
\begin{figure}[htb]
\begin{center}
\includegraphics[width=0.7\textwidth,angle=270]{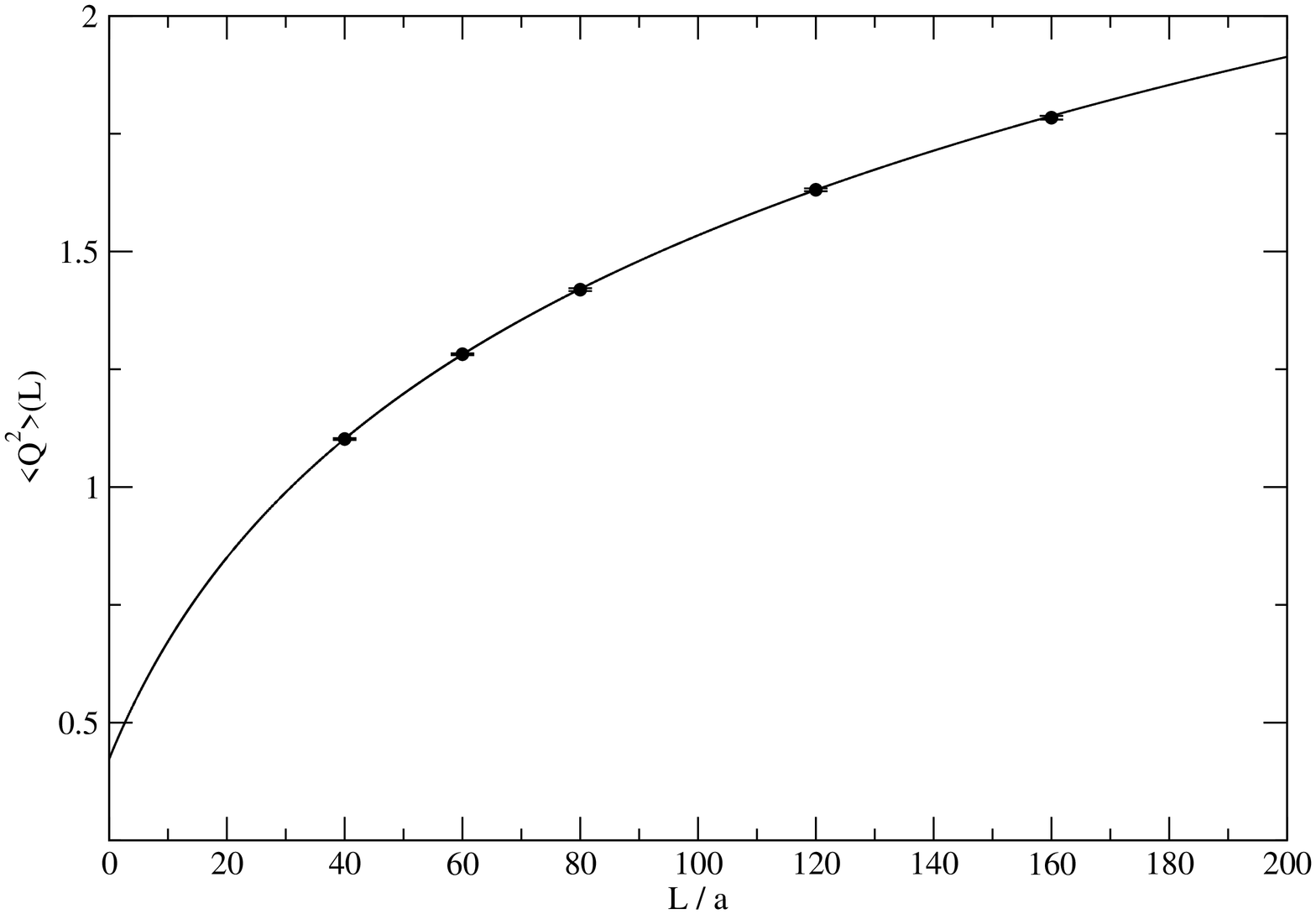}
\caption{\it The topological charge squared $\langle Q^2\rangle(L)$ at fixed 
physical size $L = 4 \xi_2(L)$, for the topological lattice action of
Subsection 4.3. The line is a fit of the form of eq.(\ref{Qfit}),  
$\langle Q^2\rangle(L) = A \log((L + L_0)/L_1)$.}
\end{center}
\end{figure}

In this case, we fit the second moment correlation length to the form
\begin{equation}
\label{xi2fit2}
\frac{\xi_2(L)}{a} = A g^2 \exp\left(\frac{2 \pi}{g^2}\right), \ 
\frac{1}{g^2} = b \lambda,
\end{equation}
which yields $A = 0.19(1)$, $b = 0.045(1)$. The results are illustrated in 
Figure 11.
\begin{figure}[htb]
\begin{center}
\includegraphics[width=0.7\textwidth,angle=270]{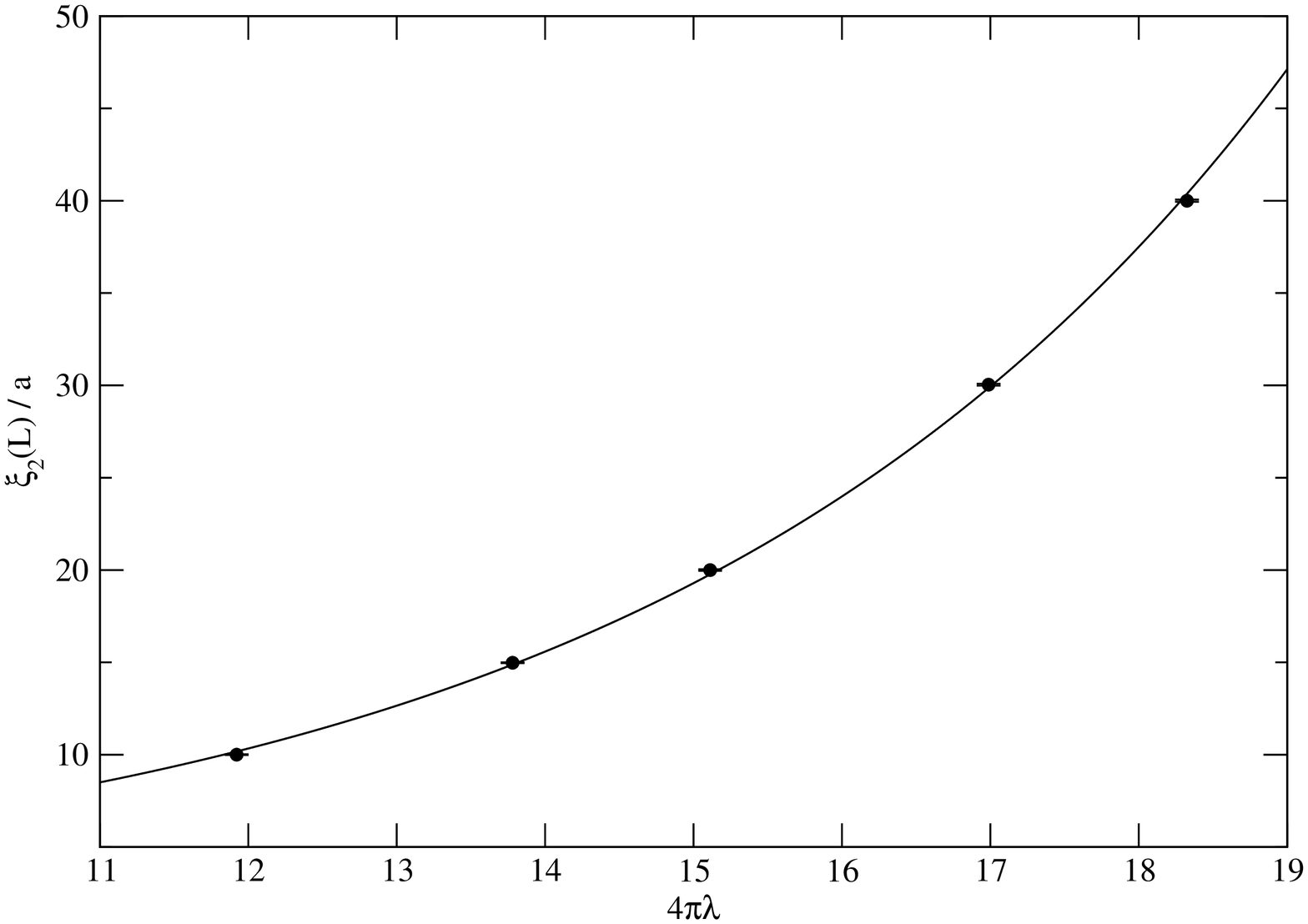}
\caption{\it Second moment correlation length $\xi_2(L)/a$ at fixed 
physical size $L = 4 \xi_2(L)$ as a function of $4 \pi \lambda$. The 
line is a fit of the form of eq.(\ref{xi2fit2}), 
$\xi_2(L)/a =  A g^2 \exp(2 \pi/g^2)$ with $1/g^2 = b \lambda$.}
\end{center}
\end{figure}
It should be pointed out that the relation between $\lambda$ and $g$ is again 
just an ansatz, which cannot be derived analytically, because perturbation 
theory is not applicable to topological lattice actions. Furthermore, one cannot
be sure that the above value of $b$ persists in the continuum limit. The lattice
Schwarz inequality (\ref{SQineq}), $S[\vec e] \geq 4 \pi \lambda |Q[\vec e]|$, 
then translates into the inequality
\begin{equation}
S[\vec e] \geq \frac{4 \pi}{b g^2} |Q[\vec e]|,
\end{equation}
which may be compared with the Schwarz inequality (\ref{Schwarz}) of the 
continuum theory $S[\vec e] \geq (4 \pi/g^2) |Q[\vec e]|$. Since $b \neq 1$, 
the lattice Schwarz inequality deviates from the one of the classical continuum 
theory. However, the lattice theory still has the correct quantum continuum 
limit.

\subsection{Correlation Function of the Topological Charge Density}

Although the topological susceptibility $\chi_t$ is logarithmically divergent 
in the 2-d $O(3)$ model, this does not imply that the whole concept of topology 
is meaningless in the continuum limit. In particular, the correlation 
function $\langle q(0) q(x) \rangle$ of the topological charge density
\begin{equation}
q(x) = \frac{1}{8 \pi} \varepsilon_{\mu\nu} \vec e(x) \cdot 
\left[\partial_\mu \vec e(x) \times \partial_\nu \vec e(x)\right],
\end{equation}
whose integral over $x$ is $\chi_t$, has a finite continuum limit for $|x| > 0$.
Analytic results for this quantity have been derived by Balog and Niedermaier
\cite{Bal97a,Bal97b}. The correlator is negative, except at $x = 0$. Up to 
logarithmic corrections, at short distances it has a power-law divergence 
proportional to $|x|^{-4}$. At $x = 0$ there is a positive divergent contact 
term. When the correlator is integrated over $x$, the numerical evidence 
obtained in the previous subsection suggests that the power-law divergence 
cancels against the contact term, but a logarithmic divergence of $\chi_t$ 
persists. As shown analytically in \cite{Giu04,Lue04}, using Ginsparg-Wilson 
quarks, also in QCD the corresponding short-distance power-law divergences
cancel against the contact terms. A corresponding study in the large $N$ limit of $\CP(N-1)$ models has been presented in \cite{Vic99}.

Let us consider the point--to--time-slice correlator 
\begin{equation}
G(x_2) = \int_0^L dx_1 \langle q(0) q(x)\rangle, \quad x = (x_1,x_2). 
\end{equation}
The corresponding quantity on the lattice receives contributions $A_{xyz}/4 \pi$
from all triangles $t_{xyz}$ in a row of plaquettes at fixed time $x_2$. Using
the meron-cluster algorithm \cite{Bie95}, we have constructed an improved
estimator for $\langle q(0) q(x)\rangle$, which receives cluster-intrinsic 
contributions only. We have measured the correlator using the 
topological angle-constraint action with $\delta/\pi = 0.4568$ at $L/a = 200$, which yields
$\xi/a = 30.9(1)$, as well as with $\delta/\pi = 0.4849$ at $L/a = 100$, which yields $\xi/a =15.54(5)$. Assuming ${\cal O}(a^2)$ cut-off effects, we have extrapolated these data to the continuum limit. In Figure 12, the extrapolated data are compared with the analytic results of \cite{Bal97a,Bal97b}.\footnote{We thank J.\ Balog for 
providing the numerical evaluation of the corresponding analytic results.}
Finite size effects are expected to be small because we have worked at $L \approx 6\xi$.
At large distances, the extrapolated Monte Carlo data agree very well with the analytic prediction, while at short distances, $x_2 / \xi < 0.35$, there are systematic deviations . We attribute these deviations to corrections to the assumed ${\cal O}(a^2)$ behavior, similar to the ones discussed before for the mass gap. In order to fully understand the cut-off effects, an analytic analysis along the lines of \cite{Bal09,Bal10}, combined with numerical data closer to the continuum limit would be most welcome. Keeping this in mind, we still conclude that our current data confirm again that the topological action leads into the
standard $O(3)$ universality class.
\begin{figure}[htb]
\begin{center}
\includegraphics[width=0.7\textwidth,angle=270]{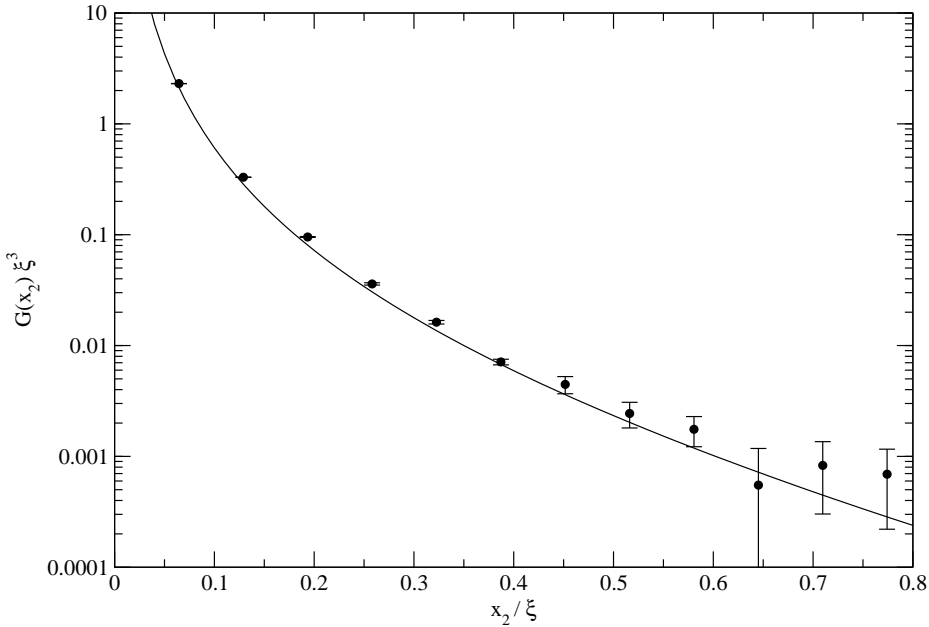}
\caption{\it The point--to--time-slice correlator 
$G(x_2) = \int_0^L dx_1 \langle q(0) q(x)\rangle$ of the topological charge 
density (in units of $m^3 = 1/\xi^3$) as a function of $x_2/\xi$. The data points are extrapolations of Monte Carlo data obtained at $\xi/a = 30.9(1)$ and $\xi/a = 15.54(5)$ to the continuum limit, assuming ${\cal O}(a^2)$ cut-off effects. At large distances, they are in good agreement with the analytic results of \cite{Bal97a,Bal97b} represented by the solid curve. We attribute the systematic deviations at short distances to not fully understood cut-off effects.}
\end{center}
\end{figure}
For the standard action a similar agreement
had already been observed in \cite{Bal97b}. In particular, this shows that some
topological quantities make perfect sense in the continuum limit of the 2-d
$O(3)$ model, despite the fact that $\chi_t$ is logarithmically divergent. The
divergence is expected to also affect the energy density of $\theta$-vacua. 
Still, other physical quantities like the $\theta$-dependent mass gap should 
have a well-defined continuum limit, which is accessible using the 
meron-cluster algorithm.

\section{Conclusions}

We have investigated topological lattice actions for the 1-d $O(2)$ and $O(3)$
as well as for the 2-d $O(3)$ model. These actions are invariant against small 
deformations of the fields. Despite the fact that topological lattice actions 
do not have the correct classical continuum limit, as we have seen, they still 
yield the correct quantum continuum limit, irrespective of whether or not they 
explicitly suppress topological charges. In particular, it does not matter
whether a topological action respects or violates a Schwarz inequality. Since, 
in contrast to other lattice actions, topological actions are invariant against 
small local deformations of the fields, one may have expected that they fall 
into a different universality class. However, the allowed local deformations 
are field-dependent and thus do not constitute a proper gauge symmetry of 
topological lattice models. In fact, even the standard action of a lattice 
$O(N)$ model has a field-dependent local $O(N-1)$ symmetry, since every spin 
can be rotated around the direction defined by the average of its nearest 
neighbors without changing the action value. Since such field-dependent local 
symmetries do not have the status of proper gauge symmetries, they have no 
impact on the corresponding universality class. 

Since topological lattice actions do not suppress small fluctuations of the 
fields, perturbation theory is not applicable. We have seen that in one 
dimension topological lattice actions suffer from strong lattice artifacts of 
${\cal O}(a)$. This seems to contradict Symanzik's effective
theory, which, however, does not apply in quantum mechanics. Despite the fact
that lattice perturbation theory cannot be applied to topological lattice 
actions, Symanzik's effective theory, which is formulated in the continuum,
still describes the lattice artifacts of the 2-d $O(3)$ model with a topological
lattice action. Interestingly, in contrast to the 1-d case, in the 2-d $O(3)$ 
model cut-off effects were observed to be less severe for a topological action 
than for the standard action, at least at practically accessible correlation
lengths. Our results may encourage the use of unconventional regularizations, 
which might be advantageous from a computational point of view.

Although the topological angle-constraint action does not explicitly suppress 
dislocations, the corresponding topological susceptibility $\chi_t$ seems not to
suffer from power-law divergences. Instead, the Monte Carlo data for $\chi_t$ 
are consistent with a logarithmic divergence, which already occurs in the
continuum theory. The numerical results for the point--to--time-slice correlator
$G(x_2) = \int_0^L dx_1 \langle q(0) q(x)\rangle$ of the topological charge 
density are consistent with the analytic predictions of 
\cite{Bal97a,Bal97b}. This underscores that, despite the fact that $\chi_t$ is 
logarithmically divergent, there exist topological physical quantities that have
a well-defined continuum limit in the 2-d $O(3)$ model. Since the corresponding 
complex action problem can be solved using the meron-cluster algorithm, a 
numerical investigation of $\theta$-vacuum effects in the 2-d $O(3)$ model is 
both feasible and physically meaningful.

It is straightforward to construct topological lattice actions for Abelian and
non-Abelian gauge theories. One may simply constrain the trace of a Wilson 
plaquette variable by some minimal value. Configurations that satisfy this 
constraint on all plaquettes can then be assigned a zero action value. It is 
conceivable that one can take algorithmic advantages from actions of this kind. 
It is an interesting subject for future studies to decide whether topological 
lattice actions may be useful in lattice Yang-Mills theory or in lattice QCD.

Our study underscores the robustness of universality, which, in particular, does
not rely on classical concepts. Even when one uses actions that do not have the 
correct classical continuum limit, cannot be treated with perturbation theory, 
or do not obey a Schwarz inequality, the emerging quantum theory still has the 
correct continuum limit. This means that the standard approach of starting from
a classical system and then quantizing it afterwards is not the only way to
define a quantum theory. Even based on concepts that make no sense 
classically, one may still be able to construct a sensible quantum theory. 
Classical physics will then emerge dynamically from the underlying quantum 
system.

\section*{Acknowledgements}

We dedicate this article to Ferenc Niedermayer on the occasion of his 65th 
birthday. Over many years, we have benefitted tremendously from his insights 
into non-perturbative physics, also in the context of this project. We gratefully
acknowledge very useful communications with J.\ Balog, P.\ Weisz, and U.\ Wolff. We also like to thank the anonymous referee for useful remarks.  W.\ B. and M.\ P. thank for the kind hospitality during visits at Bern University. This work is supported in parts by the Schweizerischer Nationalfonds (SNF). The 
``Albert Einstein Center for Fundamental Physics'' at Bern University is 
supported by the ``Innovations- und Kooperationsprojekt C-13'' of the 
Schweizerische Uni\-ver\-si\-t\"ats\-kon\-fe\-renz (SUK/CRUS).


\begin{thebibliography}{10}

\bibitem{Rei88}
T.\ Reisz, Commun.\ Math.\ Phys.\ 116 (1988) 81.

\bibitem{Rei89}
T.\ Reisz, Nucl.\ Phys.\ B318 (1989) 417.

\bibitem{Sym83}
K.\ Symanzik, Nucl.\ Phys.\ B226 (1983) 187.

\bibitem{Sym83a}
K.\ Symanzik, Nucl.\ Phys.\ B226 (1983) 205.

\bibitem{Lue85}
M.\ L\"uscher and P.\ Weisz, Commun.\ Math.\ Phys.\ 97 (1985) 59.

\bibitem{Lue85a}
M.\ L\"uscher and P.\ Weisz, Phys.\ Lett.\ B158 (1985) 250.

\bibitem{Has94}
P.\ Hasenfratz and F.\ Niedermayer, Nucl.\ Phys.\ B414 (1994) 785.

\bibitem{Bur96}
R.\ Burkhalter, Phys.\ Rev.\ D54 (1996) 4121.

\bibitem{Bur01}
R.\ Burkhalter, M.\ Imachi, Y.\ Shinno, and H.\ Yoneyama, Prog.\ Theor.\ Phys.\ 106 (2001) 613.

\bibitem{Has02}
P.\ Hasenfratz, S.\ Hauswirth, T.\ J\"org, F.\ Niedermayer, and K.\ Holland, 
Nucl.\ Phys.\ B643 (2002) 280.

\bibitem{Pat92}
A.\ Patrascioiu and E.\ Seiler, J.\ Stat.\ Phys.\ 69 (1992) 573.

\bibitem{Aiz94}
M.\ Aizenman, J.\ Stat.\ Phys.\ 77 (1994) 361.

\bibitem{Pat93}
A.\ Patrascioiu and E.\ Seiler, Nucl.\ Phys.\ Proc.\ Suppl.\ 30 (1993) 184. 

\bibitem{Pat02}
A.\ Patrascioiu and E.\ Seiler, J.\ Stat.\ Phys.\ 106 (2002) 811.

\bibitem{Has96}
M.\ Hasenbusch, Phys.\ Rev.\ D53 (1996) 3445.

\bibitem{Lue82}
M.\ L\"uscher, Commun.\ Math.\ Phys.\ 85 (1982) 29.

\bibitem{Bie95}
W.\ Bietenholz, A.\ Pochinsky, and U.-J.\ Wiese, Phys.\ Rev.\ Lett.\ 75 (1995) 
4524.

\bibitem{Her99}
P.\ Hern\'andez, K.\ Jansen, and M.\ L\"uscher, Nucl.\ Phys.\ B552 (1999) 363.

\bibitem{Lue99}
M.\ L\"uscher, Nucl.\ Phys.\ B549 (1999) 295.

\bibitem{Lue00}
M.\ L\"uscher, Nucl.\ Phys.\ B568 (2000) 162.

\bibitem{Fuk03}
H.\ Fukaya and T.\ Onogi, Phys.\ Rev.\ D68 (2003) 074503.

\bibitem{Fuk04}
H.\ Fukaya and T.\ Onogi, Phys.\ Rev.\ D70 (2004) 054508.

\bibitem{Fuk06}
H.\ Fukaya, S.\ Hashimoto, T.\ Hirohashi, K.\ Ogawa, and T.\ Onogi, 
Phys.\ Rev.\ D73 (2006) 014503.

\bibitem{Jan06}
W.\ Bietenholz, K.\ Jansen, K.-I.\ Nagai, S.\ Necco, L.\ Scorzato, and 
S.\ Shcheredin, JHEP 0603 (2006) 017.

\bibitem{Wit88}
E.\ Witten, Commun.\ Math.\ Phys.\ 117 (1988) 353.

\bibitem{DAd78}
A.\ D'Adda, M.\ L\"uscher, and P.\ Di Vecchia, Nucl.\ Phys.\ B146 (1978) 63.

\bibitem{Vic09}
E.\ Vicari and H.\ Panagopoulos, Phys.\ Rep.\ 470 (2009) 93. 

\bibitem{Ber81}
B.\ Berg and M.\ L\"uscher, Nucl.\ Phys.\ B190 (1981) 412.

\bibitem{Lue82b}
M.\ L\"uscher, Nucl.\ Phys.\ B200 (1982) 61.

\bibitem{Sch82}
P.\ Schwab, Phys.\ Lett.\ B118 (1982) 373.

\bibitem{Mue82}
G.\ M\"unster, Phys.\ Lett.\ B118 (1982) 380.

\bibitem{Lue83}
M.\ L\"uscher and D.\ Petcher, Nucl.\ Phys.\ B225 (1983) 53.

\bibitem{Bla96}
M.\ Blatter, R.\ Burkhalter, P.\ Hasenfratz, and F.\ Niedermayer, 
Phys.\ Rev.\ D53 (1996) 923.

\bibitem{Vic99}
E.\ Vicari, Nucl.\ Phys.\ B554 (1999) 301.

\bibitem{Azc07}
V.\ Azcoiti, A.\ Galante, and V.\ Laliena, Phys.\ Rev.\ Lett.\ 98 (2007) 257203.

\bibitem{Lue82a}
M.\ L\"uscher, Nucl.\ Phys.\ B205 (1982) 483.

\bibitem{Pug89}
D.\ J.\ R.\ Pugh and M.\ Teper, Phys.\ Lett.\ B324 (1989) 159.

\bibitem{Goe89}
M.\ G\"ockeler, A.\ S.\ Kronfeld, M.\ L.\ Laursen, G.\ Schierholz, and
U.-J.\ Wiese, Phys.\ Lett.\ B233 (1989) 192.

\bibitem{Phi86}
A.\ Phillips and D.\ Stone, Commun.\ Math.\ Phys.\ 103 (1985) 599.

\bibitem{Goe86}
M.\ G\"ockeler, M.\ L.\ Laursen, G.\ Schierholz, and U.-J.\ Wiese, 
Commun.\ Math.\ Phys.\ 107 (1986) 467.

\bibitem{Goe87}
M.\ G\"ockeler, A.\ S.\ Kronfeld, M.\ L.\ Laursen, G.\ Schierholz, and
U.-J.\ Wiese, Nucl.\ Phys.\ B292 (1987) 349.

\bibitem{Bie97}
W.\ Bietenholz, R.\ Brower, S.\ Chandrasekharan, and U.-J.\ Wiese, 
Phys.\ Lett.\ B407 (1997) 283.

\bibitem{Gin82}
P.\ H.\ Ginsparg and K.\ G.\ Wilson, Phys.\ Rev.\ D25 (1982) 2649.

\bibitem{Has98}
P.\ Hasenfratz, V.\ Laliena, and F.\ Niedermayer, Phys.\ Lett.\ B427 (1998) 125.

\bibitem{Giu04}
L.\ Giusti, G.\ C.\ Rossi, and M.\ Testa, Phys.\ Lett.\ B587 (2004) 157.

\bibitem{Lue04}
M.\ L\"uscher, Phys.\ Lett.\ B593 (2004) 296.

\bibitem{Wit79}
E.\ Witten, Nucl.\ Phys.\ B156 (1979) 269.

\bibitem{Ven79a}
G.\ Veneziano, Nucl.\ Phys.\ B159 (1979) 213.

\bibitem{Ven79b}
G.\ Veneziano, Phys.\ Lett.\ B95 (1980) 90.

\bibitem{Giu02}
L.\ Giusti, G.\ C.\ Rossi, M.\ Testa, and G.\ Veneziano, 
Nucl.\ Phys.\ B628 (2002) 234.

\bibitem{Deb05}
L.\ Del Debbio, L.\ Giusti and C.\ Pica, Phys.\ Rev.\ Lett.\ 94 (2005) 032003.

\bibitem{Lue10}
M.\ L\"uscher and F.\ Palombi, arXiv:1008.0732 [hep-lat]. 

\bibitem{Pol83}
A.\ M.\ Polyakov and P.\ B.\ Wiegmann, Phys.\ Lett.\ B131 (1983) 121.

\bibitem{Wie85}
P.\ B.\ Wiegmann, Phys.\ Lett.\ B152 (1985) 209.

\bibitem{Zam79}
A.\ B.\ Zamolodchikov and A.\ B.\ Zamolodchikov, Ann.\ Phys.\ 120 (1979) 
253.

\bibitem{Has90}
P.\ Hasenfratz, M.\ Maggiore, and F.\ Niedermayer, Phys.\ Lett.\ B245 
(1990) 522.

\bibitem{Bal04}
J.\ Balog und A.\ Hegedus, J.\ Phys.\ A: Math.\ Gen.\ 37 (2004) 1881.

\bibitem{Lue91}
M.\ L\"uscher, P.\ Weisz, and U.\ Wolff, Nucl.\ Phys.\ B359 (1991) 221.

\bibitem{Bal09}
J.\ Balog, F.\ Niedermayer, and P.\ Weisz, Phys.\ Lett.\ B676 (2009) 188.

\bibitem{Bal10}
J.\ Balog, F.\ Niedermayer, and P.\ Weisz, Nucl.\ Phys.\ B824(2010) 563.

\bibitem{Bal99}
J.\ Balog, M.\ Niedermaier, F.\ Niedermayer, A.\ Patrascioiu, E.\ Seiler, and 
P.\ Weisz, Phys.\ Rev.\ D60 (1999) 094508.

\bibitem{Wol89}
U.\ Wolff, Phys.\ Rev.\ Lett.\ 62 (1989) 361.

\bibitem{Wol90}
U.\ Wolff, Nucl.\ Phys.\ B334 (1990) 581.

\bibitem{Cam97}
M.\ Campostrini, A.\ Pelissetto, P.\ Rossi, and E.\ Vicari, Phys.\ Lett.\ B402 (1997) 141.

\bibitem{Car95}
S.\ Caracciolo, R.\ G.\ Edwards, A.\ Pelissetto, and A.\ D.\ Sokal,
Phys.\ Rev.\ Lett.\ 75 (1995) 1891.

\bibitem{Bea98}
B.\ B.\ Beard, R.\ J.\ Birgeneau, M.\ Greven, and U.-J.\ Wiese,
Phys.\ Rev.\ Lett.\ 80 (1998) 1742.

\bibitem{Has91}
P.\ Hasenfratz and F.\ Niedermayer, Phys.\ Lett.\ B268 (1991) 231.

\bibitem{Kro87}
A.\ S.\ Kronfeld, M.\ L.\ Laursen, G.\ Schierholz, and U.-J.\ Wiese,
Nucl.\ Phys.\ B292 (1987) 330.

\bibitem{Bal97a}
J.\ Balog and M.\ Niedermaier, Nucl.\ Phys.\ B500 (1997) 421.

\bibitem{Bal97b}
J.\ Balog and M.\ Niedermaier, Phys.\ Rev.\ Lett.\ 78 (1997) 4151.



\end{thebibliography}
\end{document}